\documentclass[aps,prl,floatfix,twocolumn,showpacs,preprintnumbers,amsmath,amssymb,superscriptaddress]{revtex4-1}
\usepackage{amsmath}
\usepackage{amssymb}
\usepackage{graphicx}
\usepackage{bm}
\usepackage{epstopdf}
\usepackage{epsfig}
\usepackage{url}

\begin{document}
\title{Spin-Lasers: Spintronics Beyond Magnetoresistance}
\author{Igor \v{Z}uti\'c}
\affiliation{Department of Physics, University at Buffalo, State University of New York, Buffalo, NY 14260, USA}
\affiliation{Department of Physics, University of Rijeka, 51000 Rijeka, Croatia}
\author{Gaofeng Xu}
\affiliation{Department of Physics, University at Buffalo, State University of New York, Buffalo, NY 14260, USA}
\author{Markus Lindemann}
\affiliation{Photonics and Terahertz Technology, Ruhr-Universit\"{a}t Bochum, Bochum, D-44780 Bochum, Germany}
\author{Paulo E. Faria Junior}
\affiliation{Institute for Theoretical Physics, University of Regensburg, 93040 Regensburg, Germany}
\author{Jeongsu Lee}
\affiliation{Institute for Theoretical Physics, University of Regensburg, 93040 Regensburg, Germany}
\author{Velimir Labinac}
\affiliation{Department of Physics, University of Rijeka, 51000 Rijeka, Croatia}
\author{Kristian Stoj\v{s}i\'c}
\affiliation{Department of Physics, University of Rijeka, 51000 Rijeka, Croatia}
\author{Guilherme M. Sipahi}
\affiliation{Instituto de F\'{i}sica de Sa\~{o} Carlos, Universidade de Sa\~{o} Paulo, 13566-590 Sa\~{o} Carlos, Sa\~{o} Paulo, Brazil}
\author{Martin R. Hofmann}
\affiliation{Photonics and Terahertz Technology, Ruhr-Universit\"{a}t Bochum, Bochum, 44780 Bochum, Germany}
\author{Nils C. Gerhardt}
\affiliation{Photonics and Terahertz Technology, Ruhr-Universit\"{a}t Bochum, Bochum, 44780 Bochum, Germany}
\begin{abstract}
Introducing spin-polarized carriers in semiconductor lasers reveals an alternative path to realize room-temperature spintronic applications, beyond the usual magnetoresistive effects. Through carrier recombination, the angular momentum of the spin-polarized carriers is transferred to photons, thus leading to the circularly polarized emitted light. The intuition for the operation of such spin-lasers can be obtained from simple bucket and harmonic oscillator models, elucidating their steady-state and dynamic response, respectively. These lasers extend the functionalities of spintronic devices and exceed the performance of conventional (spin-unpolarized) lasers, including an order of magnitude faster modulation frequency. Surprisingly, this ultrafast operation relies on a short carrier spin relaxation time and a large anisotropy of the refractive index, both viewed as detrimental in spintronics and conventional lasers. Spin-lasers provide a platform to test novel concepts in spin devices and offer progress connected to the advances in more traditional areas of spintronics.
\end{abstract}
\maketitle

\vspace{-.2cm}
\subsection{1. Introduction}
\vspace{-.2cm}

\begin{figure}[h]
\centering
\includegraphics*[width=6.4cm]{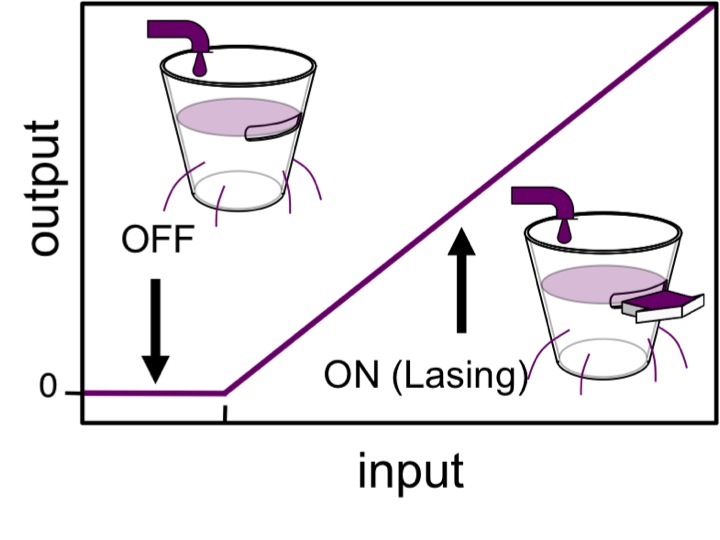}
\vspace{-0.5cm}
\caption{Light emission (output) as function of injection (input) in a conventional laser and its bucket model. 
The lasing threshold (an overfilling of a bucket) $J_T$ is marked on the axis.  From Ref.~\cite{Lee2014:APL}.
}
\label{fig:bucket}
\end{figure}

Lasers are ubiquitous devices in modern technology with applications including high-density optical storage, 
printing, medicine, optical sensing, and display systems~\cite{Chuang:2009,Coldren:2012,Yariv:1997,Michalzik:2013}. 
Fast laser are key devices in high-speed optical interconnects and the development of even faster ones is crucial due 
to the expected growth in communication and 
data centers~\cite{Michalzik:2013,Hecht2016:N,Miller2017:JLT,Miller2009:PIEEE,Hilbert2011:S,Jones2018:N}. 
One of their hallmarks is the nonlinear 
dependence of the emitted light on pumping or injection. Two regimes ({\em on} and {\em off}) are distinguished 
in Fig.~\ref{fig:bucket}. For low carrier injection 
there is only sponataneous emission, the laser operates as an ordinary light source: 
The emitted photons are incoherent. It is the higher injection
regime with stimulated emission and coherent light that makes the laser 
such a unique light source. The intensity of pumping/injection above which there is a phase-coherent emission of light is called the 
lasing threshold, $J_T$, corresponding to a kink in Fig.~\ref{fig:bucket}.

With their highly-controllable nonlinear coherent optical response, lasers have emerged
as valuable model systems to elucidate connections to other nonequilibrium and cooperative phenomena,
including  starting the field of synergetics~\cite{Haken:1985}.
The transition from an incoherent to coherent emitted light is analogous to the Landau theory of the second order 
phase transitions. In the analogy with ferromagnets, the dependence of spontaneous magnetization 
on temperature is the same as the dependence of the laser electric field on the population inversion~\cite{Haken:1985,Degiorgio1970:PRA,Degiorgio1976:PT}.

To establish an intuitive picture of conventional lasers with spin-unpolarized carriers and, subsequently, include the influence of spin polarization
in spin-lasers, we use a simple bucket model~\cite{Lee2014:APL,Lee2012:PRB}, previously considered only for conventional lasers~\cite{Parker:2005}. Water added 
to the bucket represents the injection of carriers in the laser, while the water coming out corresponds to the emitted light. 
The small holes represent carrier losses by spontaneous recombination and the large opening near the top 
depicts the lasing threshold, consistent with the operation in Fig.~\ref{fig:bucket}.

\begin{figure}[h]
\centering
\includegraphics*[width=8cm]{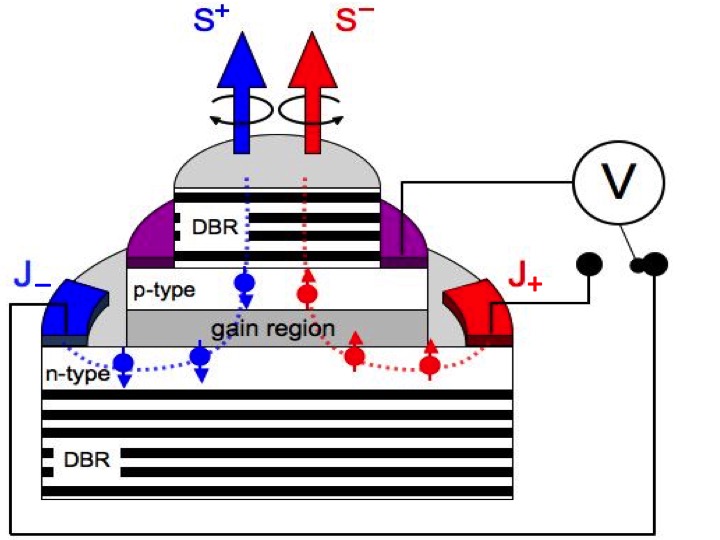}
\caption{Spin-laser scheme. 
The resonant cavity is formed by a pair of mirrors made of 
distributed Bragg reflectors (DBRs)
and the gain (active) region, typically consisting of quantum wells or dots. Electrical spin 
injection ($J_+ \neq  J_-$) is realized using two magnetic contacts. 
Spin-polarized 
carriers can also be injected optically, using circularly polarized light. The recombination of 
electrons and holes in the gain region  leads to the emission of
light of positive 
and negative helicity, $S^+$ and $S^-$. From Ref.~\cite{Sinova2012:NM}.
}
\label{fig:VCSEL}
\end{figure}

Both spin-lasers in Fig.~\ref{fig:VCSEL} and their conventional  counterparts share three main elements: 
(i) the active (gain) region, responsible for optical amplification and stimulated emission, 
(ii) the resonant cavity, and (iii) the pump, which injects (optically or electrically) energy/carriers. 
The main distinction of spin-lasers is the net carrier spin polarization (spin imbalance) in the active region, 
which leads to crucial changes in their operation. 
This spin imbalance is responsible for a circularly polarized emitted light, a result of the conservation of the 
total angular momentum during electron-hole recombination~\cite{Meier:1984,Zutic2004:RMP}. 
While carrier spin imbalance is typically lost within 1 ns or over  $<1$ $\mu$m~\cite{Zutic2004:RMP,Soldat2011:APL}, 
by being transferred to photons as circularly polarized light in spin-lasers the resulting spin-encoded signal can travel much 
faster and further, offering new paths for information transfer.

Most of the spin-lasers are implemented as 
vertical-cavity surface-emitting lasers (VCSELs)~\cite{Michalzik:2013} with the gain region based on quantum wells (QWs) or quantum dots (QDs), 
adding to their conventional counterparts electrically or optically injected spin-polarized carriers as shown 
in Fig.~\ref{fig:VCSEL}.
These lasers resemble common light-emitting diodes (LEDs) to which a resonant cavity is added. Another realization is the resonant cavity external to the 
gain device itself, using external optical elements, known as vertical-external-cavity surface-emitting lasers (VECSELs). They have 
complementary properties to VCSELs and may offer advantages for electrical spin injection~\cite{Frougier2015:OE}. 

The experimental realization of spin-lasers~\cite{Hallstein1997:PRB, Ando1998:APL,Rudolph2003:APL,Rudolph2005:APL,Gerhardt2006:EL,Holub2007:PRL,%
Hovel2008:APL,Basu2008:APL,Basu2009:PRL, Fujino2009:APL,Ikeda2009:IEEEPTL,Saha2010:PRB, Li2010:APL,Gerhardt2011:APL, %
Iba2011:APL,Frougier2013:APL, Frougier2015:OE, Hopfner2014:APL, Cheng2014:NN, Alharthi2015:APL,Hsu2015:PRB,Bhattacharya2017:PRL} 
 presents two important opportunities. The lasers provide a path to practical room-temperature spintronic devices with different 
 operating principles, not limited to magnetoresistive effects, which have enabled tremendous advances in magnetically stored 
 information~\cite{Zutic2004:RMP,Fabian2007:APS,Maekawa:2002,Parkin2004:NM,Yuasa2004:NM,Tsymbal:2019,DasSarma2000:SM,DasSarma2001:SSC}.
This requires revisiting the common understanding of material parameters for desirable operation, as well as a departure from 
more widely studied unipolar spintronic devices, where only one type of carrier (electrons) plays an active role. In contrast, 
since semiconductor lasers are bipolar devices, a simultaneous description of electrons and holes is crucial. 
On the other hand, the interest in spin-lasers is not limited to spintronics, as they could extend the limits of what is feasible with 
conventional semiconductor lasers. At room temperature an order of magnitude faster operation than  
best conventional 
lasers was demonstrated in spin-lasers~\cite{Lindemann2019:N}, while simultaneously supporting an order of magnitude 
lower-power consumption. This could enable future high-performance interconnects, which is particularly important 
since the dominant power consumption is increasingly determined by interconnects and information transfer rather than by transistors 
and information processing~\cite{Miller2017:JLT,Miller2009:PIEEE,Hilbert2011:S}.

Theoretical studies of spin-lasers mostly focus only on spin-polarized electrons, the holes are merely spectators with 
vanishingly short spin relaxation time, losing their spin polarization instantaneously. Even simple questions 
remain topics of current research. Is longer spin relaxation always better? How does a simultaneous spin relaxation of 
electrons and holes affect the operation? Could spin-lasers inspire other device concepts in spintronics?

Following this Introduction, in Section~2 we focus on the steady-state response of spin-lasers which can be intuitively understood by the analogy with a partitioned bucket and using spin-resolved rate equations. 
In Section~3 we discuss dynamic response of spin-lasers and show how many trends follow
from the analogy with a damped driven harmonic oscillator.  In Section~4 a microscopic description of
the gain region provides the connection between the electronic
structure calculations, optical gain, and birefringence. While birefringence is typically considered detrimental for 
both conventional and spin-lasers, in Section 5 we show how strong birefringence enables ultrafast operation of 
spin-lasers. We conclude with some open questions and other promising implementations of
spin-lasers which will closely rely on further developments in more common areas of spintronics.

\vspace{-.2cm}
\subsection{2. Steady-State Operation}
\vspace{-.2cm}

\subsubsection{2a. Experimental Implementation}

\begin{figure*}[t]
\begin{center}
\includegraphics*[width=1.0\linewidth]{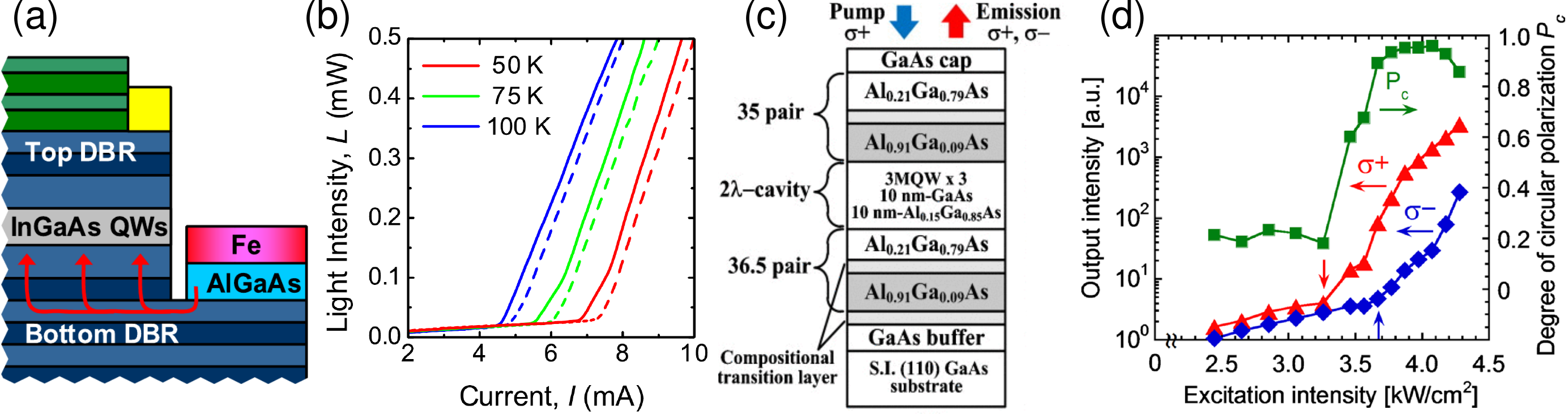}
\vspace{-0.35cm}
\caption{Experimental realizations of spin-lasers. (a) Electrical spin injection in a VCSEL using Fe injector, spin-polarized carriers
reach the In(Ga,As) QW-based gain region from Ref.~\cite{Holub2007:PRL}. (b) The solid (dashed) lines correspond to the
emitted light intensity with (without) an applied magnetic field show lasing with (without) threshold reduction from Ref.~\cite{Holub2007:PRL}.
(c) Schematic structure of a VCSEL with optical spin injection with (110) GaAs QW-based gain region from Ref.~\cite{Iba2011:APL}. (d) Emitted
helicity-resolved light intensities and optical spin polarization as a function of excitation intensity at 300 K from Ref.~\cite{Iba2011:APL}.
}
\label{fig:exp}
\end{center}
\end{figure*}

There are three major differences between spin-lasers and their conventional counterparts. 
(i)  Injected carriers are spin polarized. (ii) The light emitted 
is circularly polarized due to the spin-polarized carriers. When an electron recombines with 
a hole in accordance with optical selection rules, the electron spin orientation determines 
the helicity (circular polarization) of the emitted photon, such that the total angular momentum 
is conserved. The output polarization can be controlled by adjusting either the injection 
polarization or its intensity. (iii) There are two lasing thresholds: 
Each spin feeds 
one corresponding mode (polarization) and the imbalance of spin-up and spin-down carrier 
injection leads to two separate injection (pumping) thresholds for majority, $J_{T1}$,  and minority spin carriers, $J_{T2}$~\cite{Gothgen2008:APL,Lee2012:PRB}.

These distinguishing properties are shown in Fig.~\ref{fig:exp} for electrical~\cite{Holub2007:PRL} [(a), (b)] and 
optical injection of spin-polarized carriers~\cite{Iba2011:APL} [(c), (d)] to the QW gain region.
Optical selection rules~\cite{Meier:1984,Zutic2004:RMP} imply that only the out-of-plane component of injected spin 
contributes to the emission of circularly polarized light. Since the easy axis of the magnetization usually lies in the plane of 
the ferromagnet, electrical spin injection then requires applying magnetic field. Zero-field spin-injection
possibilities are discussed in Section~6. 
The threshold reduction in spin-lasers can be seen from Fig.~\ref{fig:exp}(b) as  the onset of lasing 
for solid lines is at smaller injection $J$, than for the broken lines. Such threshold reduction, $r$,  
is parametrized as,
\begin{equation}
r = 1 - J_{T1}/J_T,
\label{eq:r}
\end{equation}
where the majority spin threshold is less than the threshold without spin-polarized carriers, $J_{T1} < J_T$. 
Under optical excitation with positive helicity, the thresholds $J_{T1,2}$  for emitted light of positive/negative 
helicity are clearly visible in Fig.~\ref{fig:exp}(d) and marked by vertical arrows.

\subsubsection{2b. Spin-Polarized Bucket Model}
\begin{figure}[t]
\begin{center}
\includegraphics*[width=9cm]{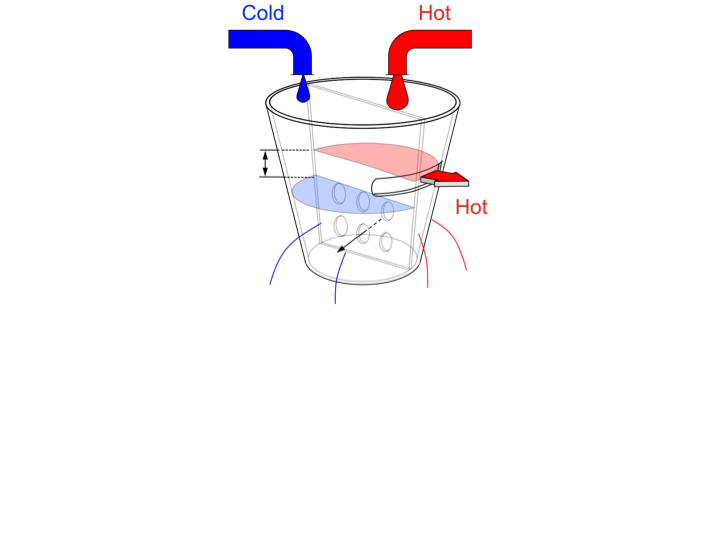}
\vspace{-3.5cm}
\caption{Spin-bucket model. The two halves denote two spin populations (hot and cold water), separately filled. Openings in the imperfect partition model the spin relaxation, 
mixing the two populations. The light emission has two lasing thresholds: 
One for each spin population.  From Ref.~\cite{Lee2012:PRB}.
}
\label{fig:sbucket}
\end{center}
\end{figure}

To understand the trends from Fig.~\ref{fig:exp} and provide an intuitive picture of spin-lasers
we invoke a bucket model shown in Fig.~\ref{fig:sbucket}. Unlike the model 
in Fig.~\ref{fig:bucket}, the bucket is partitioned in half, 
while the injection/pumping is realized from two sources of
hot and cold (spin up and down). The color code (red/blue) suggests that an equal mixture of hot and cold water, representing
injection without spin imbalance, would recover conventional spin-unpolarized (violet) behavior of laser a from Fig.~\ref{fig:bucket} as the special case.

The two halves, representing two spin populations, are separately filled with hot and cold water. 
The openings in their partition allow mixing of hot and cold water to model the spin relaxation. 
With an unequal injection of hot and cold water, the injection spin polarization is,
\begin{equation}
P_J = (J_+ - J_-)/J,
\label{eq:PJ}
\end{equation}
where the injections of the two spin projections together comprise the total injection $J = J_+  +  J_-$. 
The difference in the hot and cold water levels (Fig.~\ref{fig:sbucket}) leads to the three operating regimes and $J_{T1} \neq J_{T2}$  ($J_{T1} < J_T <J_{T2}$)~\cite{Gothgen:2010}.

At low $J$
(hot and cold water levels below the large slit), spin-up and spin-down carriers are in the off (LED) regime, thus with negligible emission. 
At higher $J$, the hot water reaches the large slit and it gushes out (Fig.~\ref{fig:sbucket}), while the amount of cold water coming out is negligible. 
The majority spin is lasing, while the minority spin is still in the LED regime;  the stimulated emission is from recombination 
of majority spin carriers. Two key consequences are  
confirmed experimentally: (a) A spin-laser will lase at a smaller $J$ than a corresponding conventional laser (only a part of the bucket needs to be filled), there is a threshold reduction, $r$ [Eq.~(\ref{eq:r}), Fig.~\ref{fig:exp}(b)]. 
(b) Even a small $P_J \ll 1$ can lead to highly circularly polarized light~\cite{Gerhardt2006:EL,Gothgen2008:APL}. This 
prediction for a ``spin-amplification"  in the 
interval $J_{T1} < J <J_{T2}$ is demonstrated in Fig.~\ref{fig:exp}(d). $P_J$ of only 4 \%
leads to the 96 \% of circuular polarization of the emitted light~\cite{Iba2011:APL}!  $J > J_{T2}$ gives rise to minority helicity photons from minority spin carriers, 
and the spin polarization of light converges to $-P_J$ with increasing injection~\cite{Gothgen2008:APL} analogous to both hot and cold water gushing out.  $J_{T2}$ as the
onset of the cold water gushing out leads to another kink in the light-injection characteristics, noticeable in solid lines for 50 K, 75 K in Fig.~\ref{fig:exp}(b). 
 
\subsubsection{2c. Spin-Polarized Rate Equations}

Phenomenological rate equations (REs),  describing the gain and the cavity region, are widely  
used in lasers~\cite{Chuang:2009,Coldren:2012}. 
We provide their spin-resolved generalization to model the spin projection and helicity of light~\cite{Lee2014:APL}. 
For example, the electron or hole density contains the spin up (+) and down (-) parts, $n= n_+ + n_-$,  $p= p_+ + p_-$,
photon density is the sum of positive and negative helicities,  $S=S^++S^-$.  Allowing unequal electron and hole injections, $J^n\neq J^p$, and spin polarizations, 
the generalized REs are, 
\begin{eqnarray}
\label{eq:REn}
dn_{\pm}/dt&=&J^n_{\pm}-g_{\pm}S^{\mp}-(n_{\pm}-n_{\mp})/\tau_{sn}-R_{sp}^{\pm}, \\
dp_{\pm}/dt&=&J^p_{\pm}-g_{\pm}S^{\mp}-(p_{\pm}-p_{\mp})/\tau_{sp}-R_{sp}^{\pm}, \\
dS^{\pm}/dt&=&\Gamma g_{\mp}S^{\pm}-S^{\pm}/\tau_{ph}+\beta \Gamma R_{sp}^{\mp}.
\label{eq:RES}
\end {eqnarray}
In the gain term, representing the stimulated emission $g_\pm(n_\pm,p_\pm,S)=
g_0(n_\pm+p_\pm-n_{\mathrm{tran}})/(1+\epsilon S)$,
$n_{\mathrm{tran}}$ is the transparency
density, $g_0$ is the gain constant, $\epsilon$ is the
gain saturation factor:  the output light $S$ does not increase indefinitely with 
$J$~\cite{Chuang:2009,Coldren:2012,Lee1993:OQE}. $\Gamma$ is the optical confinement factor. 
The electron spin relaxation is given by $(n_\pm-n_\mp) /\tau_{sn}$, 
where the electron spin relaxation time, $\tau_{sn}$, typically  $\tau_{sn} \gg \tau_{sp}$~\cite{Zutic2004:RMP,Hilton2002:PRL,Fang2017:SR}.
The carrier recombination $R_{\mathrm{sp}}^\pm$ can have various 
dependences on carrier density, i.e., linear or quadratic ($R_{sp}^+=2 Bn_+p_+$, $B$ is the bimolecular recombination coefficient)
and be characterized by a carrier recombination time $\tau_r$.
$\beta$ is the fraction of the spontaneous recombination producing
light  coupled to the resonant cavity, $\tau_{\mathrm ph}$ is the
photon lifetime, to model optical losses~\cite{Gothgen2008:APL,Holub2007:PRL}. 
For an LED, the gain term vanishes and $\beta=1$, while for VCSELs, $\beta \ll 1$.

Before showing how these REs reveal surprising trends in the dynamic operation of spin-lasers (Section~3)
and complementing their description by a microscopic analysis of the gain term (Section~4), we use them to corroborate
the picture for the steady-state operation of spin-lasers, as expected from the bucket model.

The threshold reduction is readily obtained from the steady-state solution of Eqs.~(\ref{eq:REn})$-$(\ref{eq:RES}). 
With quadratic recombination and $P_J=0$, $J_T=B n_T^2\equiv n_T/\tau_r$, where we use threshold density for
carriers, $n_T =(\Gamma g_0 \tau_{\mathrm{ph}})^{-1}+n_{\mathrm{tran}}$, and photons, $S_T=J_T \Gamma \tau_{ph}$.
For $\tau_{sn}=\tau_{sp}$, $P_J=1$ and in the limit $\tau_r/\tau_{sn} \ll 1$, $J_{T1}=J_T/2$ or, equivalently, $r=1/2$, recall Eq.~(\ref{eq:r}). 
As expected from the bucket model filled only with hot water and the perfect partition, we only need to fill a half of the bucket for it to overflow. 
In the opposite limit, $\tau_r/\tau_{sn} \gg 1$, 
$r=0$, there is no threshold reduction since the bucket partition is negligible (fast spin relaxation),
we need to fill the whole bucket.  In spin-lasers the electron-hole symmetry is broken, typically $\tau_{sn} \gg \tau_{sp}$, 
with the resulting threshold~\cite{Gothgen2008:APL},  
\begin{equation}
r=1-\frac{ 
\left[ 1-2\tau_r/\tau_{sn}+\sqrt{1+ 4\tau_r/\tau_{sn}
(1+\tau_r/\tau_{sn} + |P_J|)} \right]^2
}
{\left( 2+ |P_J| \right)^2},
\label{eq:QR}
\end{equation}
 For $P_J=1$ and $\tau_r/\tau_{sn}  \gg 1$, again $r=0$. Surprisingly, for $\tau_r/\tau_{sn}  \ll 1$,  $r=5/9$,  
more than expected from Fig.~\ref{fig:sbucket} and previously thought maximum 
$r= 1/2$~\cite{Rudolph2003:APL,Holub2007:PRL,Oestreich2005:SM}.

\subsection{3. Dynamic Operation}

\subsubsection{3a. Harmonic Oscillator Model}

\begin{figure}[h]
\centering
\includegraphics*[width=8cm]{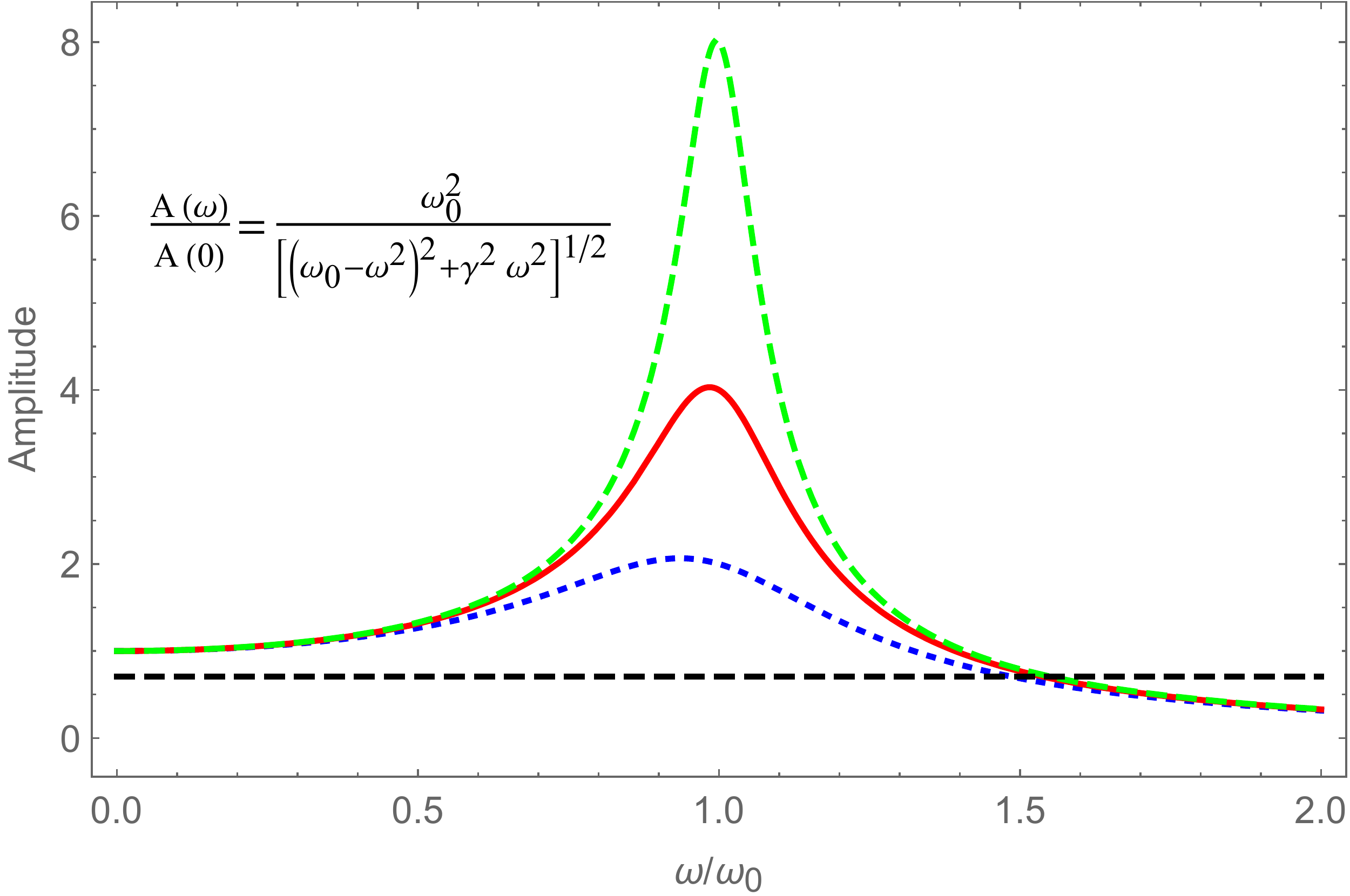}
\caption{Normalized amplitude,  $A(\omega)/A(0)$, of a driven harmonic oscillator.
The curves are given for damping constants $\gamma=\omega_0/2, \omega_0/4, \omega_0/8$, where $\omega_0$
is the angular frequency of the simple harmonic oscillator. Horizontal line: -3 dB signal. 
}
\label{fig:HO}
\end{figure}

A  damped driven harmonic oscillator, ${\ddot x}+\gamma {\dot x}+\omega^2_0x=(F_0/m) \cos\omega t$,
provides a valuable model for the dynamic operation of lasers~\cite{Lee2012:PRB}, which also share its resonant behavior near the angular frequency 
$\omega \approx \omega_0$ and a large reduction of the amplitude, $A(\omega)$, for $\omega \gg \omega_0$, see Fig.~\ref{fig:HO},
\begin{equation}
A(\omega)/A(0)=\omega_0^2/\left[(\omega_0^2-\omega^2)^2+\gamma^2\omega^2\right]^{1/2},
\label{eq:HO}
\end{equation}
where $\omega_0$ is the  angular frequency of the simple harmonic 
oscillator, $\gamma$ is the damping constant, $F_0$ is the amplitude 
of the driving force and $m$ is the mass. As we will later show,  $A(\omega)$ reduction from 
$A(0)$ by -3 dB (Fig.~\ref{fig:HO}), gives a frequency range for a still
substantial signal, corresponding to the modulation bandwidth of a laser~\cite{Chuang:2009,Lee2012:PRB,Xu2020:P}. 

\subsubsection{3b. Small Signal Analysis}
\label{sec:SSA}

\begin{figure*}[t]
\centering
\includegraphics*[width=0.75\linewidth]{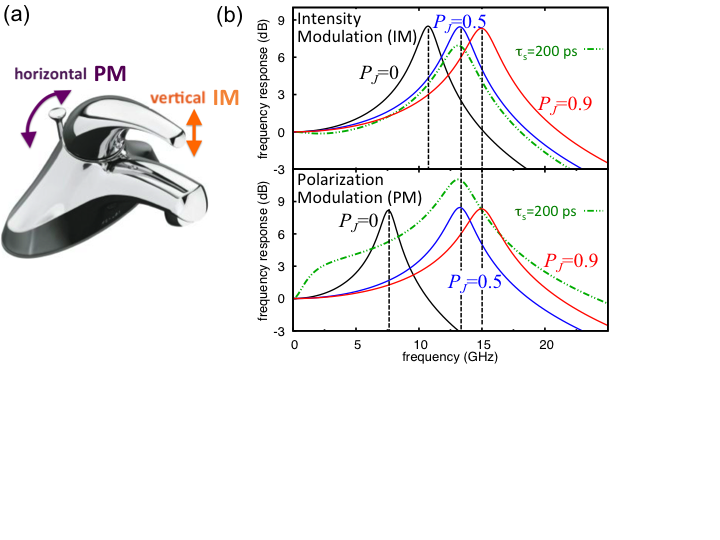}
\vspace{-3.2cm}
\caption{(a) Faucet analogy with intensity/polarization modulation (IM, PM) of a laser. (b) Small signal analysis of {\em IM}
($\delta J/J_0=0.01$) and {\em PM}
($\delta P_J=0.01$), at $J_0=1.9 J_T$.
The frequency response function $|\delta S^-(\omega)/\delta J_+(\omega)|$ is
normalized its $\omega=0$ value. Results are
shown  in the  limit of $1/\tau_{sn}=\beta=\epsilon=0$,
except broken lines for $\tau_{sn}=\tau_r$, $\beta=10^{-4}$, and
$\epsilon=2\times 10^{-18}$ cm$^3$. The frequency response at $-3$ dB value
gives the bandwidth of the laser~\cite{Chuang:2009}.
The vertical lines denote approximate
peak positions evaluated at $1/\tau_{sn}=\epsilon=\beta=0$ from
Eq.~(\ref{eq:om2}), except for {\em PM} at $P_{J0}=0$ using Eq.~(\ref{eq:om3}). When $P_{J0}=0$
there is a small difference between $\epsilon=\beta=0$ and finite $\epsilon, \beta$
results for {\em IM}, the difference is nearly invisible for {\em PM}. From Ref.~\cite{Lee2010:APL}.
}
\label{fig:SSA}
\end{figure*}

The most attractive properties of conventional lasers are in their dynamic performance~\cite{Chuang:2009}.
Here we explore spin-polarized modulation from REs (\ref{eq:REn})$-$(\ref{eq:RES}).
The model of damped driven harmonic oscillator in Section~3b is particularly useful for dynamic operation of 
VCSELs (recall Figs.~\ref{fig:VCSEL}, \ref{fig:exp}), the most common implementations of spin-lasers on which we focus 
here. External-cavity lasers (VECSELs)~\cite{Alouini2018:OE} and quantum cascade lasers~\cite{Faist:2013}
correspond to an overdamped oscillator and thus have no significant resonant behavior.  

The intuition from the harmonic oscillator model gives a simple principle that lasers, as many other systems,
have a frequency range over which they can follow externally imposed perturbation. For higher frequencies the 
laser response significantly diminishes compared to its zero-frequency response, just as we see in Fig.~\ref{fig:HO}
that $A(\omega \gg \omega_0) \rightarrow 0$, and the resulting low signal to noise ratio is too small and thus limits the 
frequency range over which there could be a useful information transfer. Commonly, such a useful frequency range is given by the -3 dB 
signal reduction (see Fig.~\ref{fig:HO}) and referred to as the modulation bandwidth~\cite{Chuang:2009,Lee2010:APL}.

Therefore, achieving higher modulation 
frequencies could enable an enhanced modulation bandwidth of a laser and thus enhanced information transfer. In the following
analysis we examine if spin-polarized carriers may offer such higher modulation frequencies.  Each of the key quantities, 
$X$ (such as, $J$, $S$, $n$ and $P_J$), can be decomposed into a steady-state $X_0$ and a modulated part
$\delta X(t)$, $X=X_0+\delta X(t)$.

We focus  on the  intensity and polarization modulation
({\em IM}, {\em PM}), illustrated in Fig.~\ref{fig:SSA}.
{\em IM} for a steady-state polarization implies $J_+ \neq J_-$
(unless $P_J=0$),
\begin{equation}
IM: \:
J=J_0+\delta J \cos(\omega t), \quad  P_J = P_{J0},
\label{eq:AM}
\end{equation}
where $\omega$ is the angular modulation frequency.
Such a modulation can be contrasted with {\em PM}
which also has $J_+ \neq J_-$, but
$J$ remains constant~\cite{const},
\begin{equation}
PM: \:
J=J_0, \quad P_J=P_{J0}+\delta P_J \cos(\omega t).
\label{eq:PM}
\end{equation}

To analyze the dynamic operation of the laser a perturbative approach to the steady-state response is commonly used
and referred to as the small signal analysis (SSA), limited to a small modulation
($|\delta J/J_0| \ll 1$ for {\em IM} and $|\delta P_J| \ll 1,  |P_{J0}\pm \delta P_J|<1$ for {\em PM}.
From REs we can obtain $\delta S^{\pm}(\omega)$ and the generalized (modulation) frequency response functions 
$R_\pm(\omega)=|\delta S^\mp(\omega)/\delta J_\pm(\omega)|$. In the $P_J=0$ limit they reduce to, 
$R(\omega)=|\delta S(\omega)/\delta J(\omega)|$, usually normalized to its $\omega=0$ value, just as in Eq.~(\ref{eq:HO}),
\begin{equation}
\left|  R(\omega)/R(0) \right| = \omega_R^2/\left[(\omega_R^2-\omega^2)^2+\gamma^2\omega^2\right]^{1/2},
\label{eq:band}
\end{equation}
where, 
$\omega_R^2 \approx  g_0 S_0/[\tau_{ph}(1+\epsilon S_0)]$,
is the square of the (angular) relaxation oscillation frequency and  
$\gamma \approx 1/\tau_r+ (\tau_{ph}+\epsilon/g_0)\omega_R^2$ is the damping factor.

From SSA with a linear recombination, $(n-n_0)/\tau_r$, $\epsilon=0$, and for $P_{J0}=0$ and $J>J_{T2}$,  we obtain~\cite{Lee2010:APL}, 
\begin{eqnarray}
\omega^2_{R, IM} \approx 2\omega^2_{R, PM} \approx
\Gamma g_0 (n_T/\tau_r)(J_0/J_T-1), 
\label{eq:om3}
\end{eqnarray}
where we write $\omega^2_{R, IM}= g_0 S_0/\tau_{ph}$ to recover the standard result of conventional lasers~\cite{Chuang:2009}.
For  a  steady-state spin injection  $P_{J0}\neq0$ and  $J_{T1}<J<J_{T2}$,  $1/\tau_{sn}=0$,
\begin{eqnarray}
\omega^2_{R, IM}=\omega^2_{R, PM} &\approx&
\Gamma g_0 (n_T/\tau_r)\left[(1+|P_{J0}|/2)J_0/J_T-1\right]  \nonumber \\
&=&3g_0 S_0^-/2\tau_{\mathrm{ph}}.
\label{eq:om2}
\end{eqnarray}
The trends from Eqs.~(\ref{eq:om3}), (\ref{eq:om2}) are shown in Fig.~\ref{fig:SSA},
with spin injection $\omega_R$ and the bandwidth exceed those in conventional lasers. 
$P_{J0}=0$ (using finite $\epsilon$ and $\beta$~\cite{Rudolph2005:APL,Holub2007:PRL}) 
show that our analytical approximations
for $\epsilon=\beta=0$ are accurate at moderate $J$.
The increase of $\omega_R$ and the bandwidth with $P_{J0}$,  for {\em IM}
and {\em PM}, can be understood as the dynamic manifestation of threshold
reduction with increasing $P_{J0}$.
With $\omega_R \propto (S_0^-)^{1/2}$ [Eq.~(\ref{eq:om2})],
the situation is analogous to the
conventional lasers: $\omega_R$ and the bandwidth both increase with
the square root of the output power~\cite{Chuang:2009,Yariv:1997}
($S_0^+=0$ for $J_{T1}<J<J_{T2}$).

An important advantage of spin-lasers is that the increase in $S_0^-$ can be achieved even at
{\em constant} input power (i.e., $J-J_T$), simply by increasing $P_{J0}$.
A larger $P_{J0}$ allows for a larger  $J_0$ (maintaining $J_{T1}<J_0<J_{T2}$),
which can further enhance the bandwidth, as seen in Eq.~(\ref{eq:om2}).
For $P_{J0}=0.9$, $J_0$ can be up to $10J_T$~\cite{Gothgen2008:APL}.

We next examine the effects of $\tau_{sn} \neq 0$, shown for $\tau_{sn}=\tau_r$,
and $P_{J0}=0.5$.  {\em IM} follows a plausible trend:
$\omega_R$ and the bandwidth monotonically decrease and eventually attain
``conventional'' values for $\tau_{sn} \rightarrow 0$.
Surprisingly,  a seemingly detrimental spin
relaxation {\em enhances} the {\em PM} bandwidth and the peak in the frequency response,
as compared to the long $\tau_{sn}$  ($\tau_{sn}\gg \tau_r$) limit~\cite{Lee2010:APL,Banerjee2011:JAP}.
A shorter $\tau_{sn}$ reduces $P_J$ and thus the amplitude of modulated light.
Since $\delta S^-(0)$ decreases faster with $\tau_r/\tau_{sn}$
than $\delta S^-(\omega>0)$, we find
an increase in the normalized response function, shown
in Fig.~\ref{fig:SSA}.
The bandwidth increase comes at the cost of a reduced modulation signal.

The above trends allow us to infer some other possible advantages of {\em PM} at fixed
injection. For example, such spin-lasers could reduce parasitic frequency changes, so-called chirp 
($\alpha$-factor)~\cite{Yariv:1997}, associated with the {\em IM} since $\delta J$ leads to  $\delta n(t)$ 
and, as expected from Kramers-Kroning relations, dynamically change the refractive index which influences the frequency of the emitted light~\cite{Boeris2012:APL}.

\subsubsection{3c. Large Signal Analysis}

We next turn to the large-signal analysis~\cite{Lee2014:APL} for abrupt
changes between {\em off} and {\em on} injection, $J_\mathrm{off}$, $J_\mathrm{on}$, important for high modulation 
frequency, short optical pulse generation,  and high bit-rate in 
telecommunication~\cite{Coldren:2012,Petermann:1988}.
In usually employed GaAs- or InAs-based QWs, the spin relaxation times for holes are negligible. For example,
at 300 K in GaAs-based QWs $\tau_{sp} \sim 1-2$ ps, while $\tau_{sn} \sim 80-120$ ps~\cite{Fang2017:SR}, consistent
with our focus on the limit $\tau_{sp}/\tau_{sn} \ll 1$.
However, our generalization to explicitly consider spin polarization of holes in RE~(4)~\cite{Lee2014:APL,Wasner2015:APL}
and optical gain calculation~\cite{FariaJunior2017:PRB} (see Section~4)
could be helpful for less studied gain regions  based on QDs, GaN~\cite{Cheng2014:NN,FariaJunior2017:PRB},
or transition metal dichalcogenides~\cite{Lee2014:APL,Mak2012:NN}.

\begin{figure}[h]
\centering
\includegraphics*[width=14.5cm]{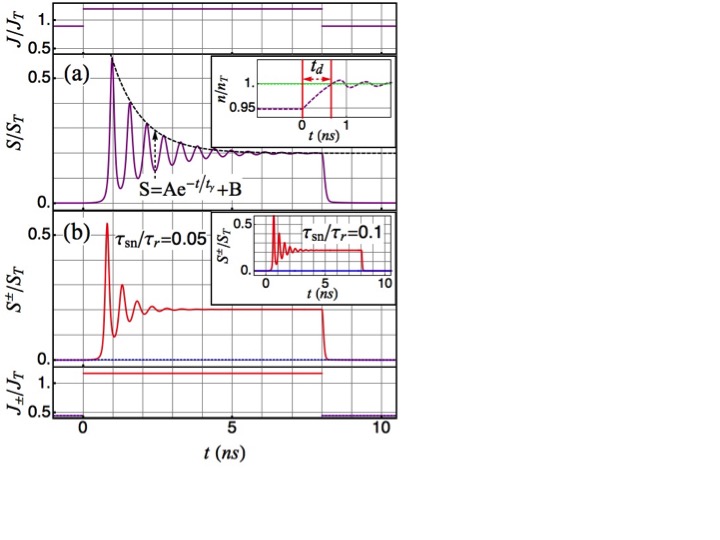}
\vspace{-2cm}
\caption{Large signal modulation. Time-dependence of {\em AM} of (a) injection, emitted 
light, and (inset) carrier density for a conventional laser, all normalized to $P_J=0$ threshold values.
Delay and decay times, $t_d$ and $t_\gamma$.
(b) {\em PM} of injection, emitted light,
for a spin-laser with $\tau_{sn}/\tau_r=0.05$,  and (inset) 
emitted light  for $\tau_{sn}/\tau_r=0.1$. 
$P_J^n=1$, $\tau_{sp}=0$. 
From Ref.~\cite{Lee2014:APL}.
}
\label{fig:LSA}
\end{figure}

In conventional lasers a step increase in injection leads to
a turn-on delay time, $t_\mathrm{d}$, required for carrier density to build up to the threshold value before
light is emitted, generally detrimental for the high-speed communication.
Such delay can be obtained for $\beta=0$ [see Eq.~(\ref{eq:RES})] and
$J_{\mathrm{off}} <J_T <J_{\mathrm{on}}$ by~\cite{Petermann:1988}, for the linear recombination,
\begin{equation}
t_d=\tau_r \ln \left[ (J_{\mathrm{on}}-J_{\mathrm{off}} )/(J_{\mathrm{on}}-J_T) \right]. 
\label{eq:tdLR}  
\end{equation}
For the quadratic recombination~\cite{Coldren:2012},
\begin{equation}
t_d=\tau_r \sqrt{\frac{J_T}
{J_\mathrm{on}}}
\bigg[\!\tanh^{\!-1}\! \! \sqrt{J_T/J_\mathrm{on}}
-\tanh^{\!-1}\! \!\sqrt{J_\mathrm{off}/J_\mathrm{on}}
\bigg],
\label{eq:tdQR}  
\end{equation}
where  $\tau_r=1/(Bn_T)$.
Another turn-on characteristic of the large-signal modulation is an overshoot
of the emitted light and its damped ringing, often undesirable
and implying distortion for a pulsed laser operation~\cite{Coldren:2012,Petermann:1988}.

Could these turn-on characteristics be improved in spin-lasers, and what is the 
corresponding effect of spin relaxation time? We first examine the maximum
asymmetry between the $\tau_{sn}>0$ and $\tau_{sp}=0$, 
for optimal threshold reduction $r$, recall Eq.~(\ref{eq:r}). 
Our results in Fig.~\ref{fig:LSA} provide  a direct comparison between 
large-signal modulation in conventional and spin-lasers having  the same 
{\em on}-state
photon density after the end of turn-on transients. From Eqs.~(\ref{eq:tdLR}) 
or (\ref{eq:tdQR}) we can also infer trends for spin-lasers: an increase in $\tau_{sn}$ 
leads to the threshold reduction and thus the monotonic decrease of $t_d$.
These equations also apply for spin-lasers by replacing 
$J_T$ with $J_{T1}$,  as long as $J_{\mathrm{off}} <J_{T1} <J_{\mathrm{on}}$.
For a conventional laser in Fig.~\ref{fig:LSA}(a), $t_d=0.64$ ns from Eq.~(\ref{eq:tdQR})  ($\beta=0$), 
or $t_d=0.66$ ns  ($\beta=1.8\times 10^{-4}$) obtained numerically.
The delay for a spin-laser is reduced in Fig.~\ref{fig:LSA}(b)
 to  $t_d=0.53$ ns,  even for a short spin relaxation time, 
$\tau_{sn}/\tau_r=0.05$, and to  $t_d=0.36$ ns for $\tau_{sn}/\tau_r=0.1$ (inset). 

The step-function injection causes an abrupt increase and overshoot in the photon and carrier 
densities. Through damped oscillations,  their densities relax to the steady-state values.  
From ringing patterns in Figs.~\ref{fig:LSA}(a) and (b) we can estimate: (i)  a modulation
bandwidth from the corresponding relaxation oscillation frequency, 
$\omega_R\propto$ 1/period of damped oscillations~\cite{Petermann:1988},
and (ii) the decay time of the damped oscillations, $t_\gamma$.
The ringing pattern confirms another advantage of spin-lasers:  enhanced 
$\omega_R$, as compared to conventional lasers (recovered for $\tau_{sn},\tau_{sp}\rightarrow 0$). 
In Section~3b, for linear recombination, we have seen that $\omega_R$ is enhanced with the threshold 
reduction~\cite{Lee2010:APL}. 
For quadratic recombination from SSA with $P_J=1$, $\tau_{sn}\rightarrow \infty$,  $J_{T1}< J_T< J_{T2}$, 
we obtain~\cite{Lee2014:APL}, 
\begin{equation}
\omega_R^2=\Gamma g_0 [(3/2)J-(2/3)J_T],  \quad 
\omega_R^2=\Gamma g_0 [2J-J_T],
\label{eq:OmegaR}  
\end{equation}
for $\tau_{sp}=0$  and $\tau_{sp}\rightarrow \infty$, respectively.  
These results match well the numerically extracted $\omega_R$ from a 
ringing pattern and confirm $> 100$ \% bandwidth enhancement in spin-lasers.
To maximize the bandwidth enhancement, comparable $\tau_{sn}$ and $\tau_{sp}$ are desirable, 
unlike when maximizing $r$, where their asymmetry is required~\cite{Gothgen2008:APL,Lee2014:APL}. 

Enhancing damping in the ringing pattern is desirable for switching lasers with 
well-defined $J_\mathrm{on}$, $J_\mathrm{off}$ values. In conventional lasers near $J_T$  
such damping is  dominated by the recombination processes and photon decay. A parametrization of this damping 
by $t_\gamma$ [Fig.~\ref{fig:LSA}(a)] reveals a peculiar behavior in spin-lasers. 
Let us focus on the $\tau_{sp}=0$ case. 
 The damping is enhanced by  the spin relaxation due to $n_+ \neq n_-$:
a shorter $\tau_{sn}$ provides a stronger damping.
However, spin-lasers reduce to conventional lasers for  $\tau_{sn} \to 0$; the injected 
spin polarization is immediately lost and the damping mechanism from spin relaxation 
is suppressed. Comparing then Fig.~\ref{fig:LSA}(a) [viewed as the spin-laser with
$\tau_{sn} \to 0$] with Fig.~\ref{fig:LSA} (b) and its inset confirms the {\em nonmonotonic} damping of  ringing. 
 
This surprising nonmonotonic  dependence of $t_\gamma(\tau_{sn})$ can be explained in analogy 
with Matthiessen's rule for different scattering mechanisms~\cite{Lee2014:APL}.
It is instructive to view the effective decay as an interplay of the spin-independent and 
spin-dependent part: $1/t_\gamma=1/t_\gamma^0+1/t_\gamma^{sn}$.
Similarly, it is customary to express~\cite{Coldren:2012} the carrier lifetime $t_c$ as a 
combination of radiative (R) and nonradiative (NR) contributions: $1/t_c=1/t_c^R+1/t_c^{NR}$.
In the limit of $\tau_{sn} \rightarrow 0$, the spin contribution vanishes ($1/t_\gamma^{sn} \rightarrow 0$)
and $t_\gamma=t^0_\gamma \approx 1$ ns, as in the conventional lasers~\cite{Lee2014:APL}.
For the maximum decay  the two contributions should be comparable: 
$1/t^{\rm{MIN}}_\gamma=2/t_\gamma^0$. Turning now  to the case of $\tau_{sn}=\tau_{sp}$, we infer an additional 
hole-spin contribution  $1/t^{sp}_\gamma$,  which leads to even smaller $t^{\rm{MIN}}_\gamma=t_\gamma^0/3$.  

Our laser analysis with the explicit inclusion of spin relaxation times 
for both electrons and holes reveals that optimizing the performance in spintronic devices
is much more complex than simply requiring suppressed spin relaxation. Since
the spin relaxation times for a given material could be readily changed by an applied magnetic 
and even electric field~\cite{Zutic2004:RMP}, it is possible to test some of our predicted trends 
in already fabricated spin-lasers. 

\subsection{4. Microscopic Description}

To better understand the operation of spin-lasers and how their performance could exceed
that of best conventional lasers it is important to complement their RE description with a microscopic
picture. We focus on the gain region which usually includes III-V QWs or QDs. At the
level of REs, there is an effective mapping~\cite{Lee2012:PRB,Oszwaldowski2010:PRB},
 between QW- and QD-based lasers~\cite{Basu2008:APL,Basu2009:PRL,Saha2010:PRB}.

\subsubsection{4a. Optical Gain}
\label{OG}

\begin{figure}[h]
\includegraphics*[width=7.5cm]{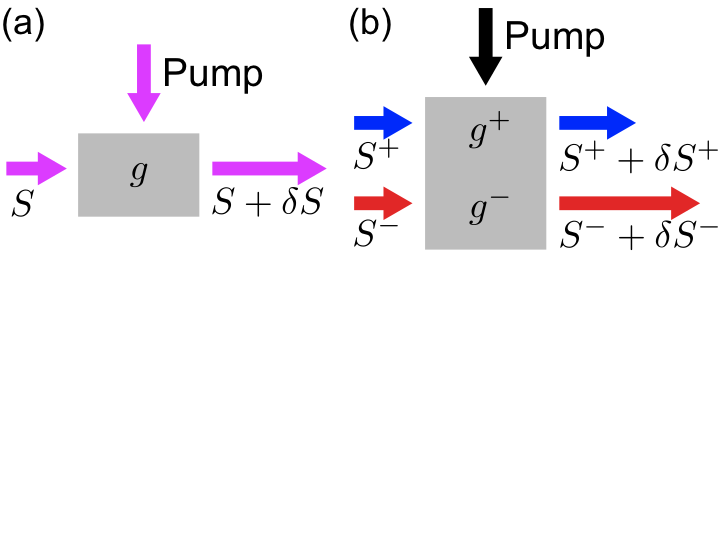}
\vspace{-3.15cm}
\caption{Schematic of the optical gain, $g$,  for (a) conventional and (b) spin-laser. 
With pumping/injection, a photon density $S$ increases by $\delta S$ 
as it passes across the gain region. In the spin-laser this increase depends on 
the positive ($+$)/negative($-$) helicity of the light. 
From Ref.~\cite{FariaJunior2015:PRB}.
}
\label{fig:gain}
\end{figure}

The key effect of the gain region is to produce a stimulated emission and coherent 
light that makes the laser such a unique light source. The corresponding optical 
gain that describes stimulated emission, under sufficiently strong pumping/injection 
of carriers, can be illustrated in Fig.~\ref{fig:gain}.
Neglecting any loses in the resonant cavity, such a gain provides an 
exponential growth rate with the distance across a small segment of gain 
material~\cite{Coldren:2012}. Since both static and dynamic operations of spin-lasers 
crucially depend on their corresponding optical gain, we consider 
its microscopic description derived from an accurate electronic structure of an 
active region.

\subsubsection{4b. Theoretical Framework}
\label{theory}

We focus  here on the QW implementation also  found 
in most of the commercial VCSELs~\cite{Michalzik:2013}. To obtain an accurate electronic 
structure in the active region, needed to calculate optical gain, we use the $8 \times 8$ 
$\bm{k{\cdot}p}$ 
method~\cite{Cardona:2010,FariaJunior2015:PRB}.
Previous calculations were performed using  $6 \times 6$ 
$\bm{k{\cdot}p}$ method in which the electronic structure
for conduction and valence bands is assumed decoupled~\cite{Holub2011:PRB}. 
The total Hamiltonian of the QW system, with 
the growth axis along the $z$ direction, can be written as,
\begin{equation}
H_{\textrm{QW}}(z) = H_{\textrm{kp}}(z) + H_{\textrm{st}}(z) + H_{\textrm{O}}(z),
\label{eq:HQW}
\end{equation} 
where $H_{\textrm{kp}}(z)$ denotes the $\bm{k{\cdot}p}$ 
 term, $H_{\textrm{st}}(z)$ describes the 
strain term, and $H_{\textrm{O}}(z)$ includes the band-offset at the interface 
that generates the QW energy profile.

Considering that common nonmagnetic semiconductors are well characterized by the 
vacuum permeability, $\mu_0$, a complex dielectric function $\varepsilon(\omega)=\varepsilon_r(\omega)+\varepsilon_i(\omega)$, 
where $\omega$ is the photon (angular) frequency, can be used to simply express 
the dispersion and absorption of electromagnetic waves. The absorption coefficient 
describing gain or loss of the amplitude of an electromagnetic wave propagating 
in such a medium is the negative value of the gain coefficient (or gain 
spectrum)~\cite{Chuang:2009}. This gain coefficient corresponds to the ratio of the number of 
photons emitted per second per unit volume and the number of injected photons per second per unit area, 
therefore having a dimension of 1/length, it can be expressed as,
\begin{equation}
g^a(\omega)=-\frac{\omega}{c n_r}
\epsilon^a_i(\omega) \, ,
\label{eq:gain}
\end{equation}
where $c$ is the speed of light, $n_r$ is the dominant real part of the refractive 
index of the material~\cite{Haug:2004} and $\varepsilon^a_i(\omega)$ is the imaginary 
part of the dielectric function which generally depends on the polarization of 
light, $a$, given by
\begin{equation}
\varepsilon^a_i(\omega)=C_0{\sum}_{c,v,\vec{k}} \left| p^a_{cv\vec{k}} \right|^2 \left(f_{v\vec{k}}-f_{c\vec{k}}\right) 
\delta\left[\hbar\omega_{{cv}\vec{k}}-\hbar\omega\right], 
\label{eq:epsI}
\end{equation}
where the indices $c$ (not to be confused with the speed of light) and $v$ label 
the conduction and valence subbands, respectively, $\vec{k}$ is the wave vector, 
$p^a_{cv\vec{k}}$ is the interband dipole transition amplitude 
which is directly calculated using $\bm{k{\cdot}p}$ wave functions,
$f_{c(v)\vec{k}}$ 
is the Fermi-Dirac distribution for the electron occupancy in the conduction (valence) 
subbands, $\hbar$ is the Planck's constant, $\omega_{{cv}\vec{k}}$ is the interband 
transition frequency, and $\delta$ is the Dirac delta-function, which is often 
replaced to include broadening effects for finite lifetimes~\cite{Chuang:2009,Chow:1999}.
From the previous studies of conventional lasers, excluding spin injection, such broadening 
effects are known to improve~\cite{Chow:1999} the free-carrier gain model we employ here. For a full 
microscopic  description and quantitative accuracy, one would also need to include 
many-body effects~\cite{Chow:1999} and generalize them to spin-lasers.
The constant $C_0$ is $C_0 = 4\pi^2 e^2/(\varepsilon_0 m_0^2\omega^2\Omega)$,
where $e$ is the electron charge, $m_0$ is the free electron mass, and $\Omega$ 
is the QW volume.

Using the dipole selection rules for the spin-conserving interband transitions, 
the gain spectrum, 
\begin{equation}
g^a(\omega) = g^a_+(\omega) + g^a_-(\omega)
\label{eq:gain_sum}
\end{equation}
can be expressed in terms of the contributions of spin up and down carriers. To 
obtain $g^a_{+(-)}(\omega)$, the summation of conduction and valence subbands is 
restricted to only one spin: ${\sum}_c \rightarrow {\sum}_{c+(-)}$ 
and ${\sum}_v \rightarrow {\sum}_{v+(-)}$ in Eq.~(\ref{eq:epsI}).

\begin{figure}[h]
\begin{center}
\includegraphics{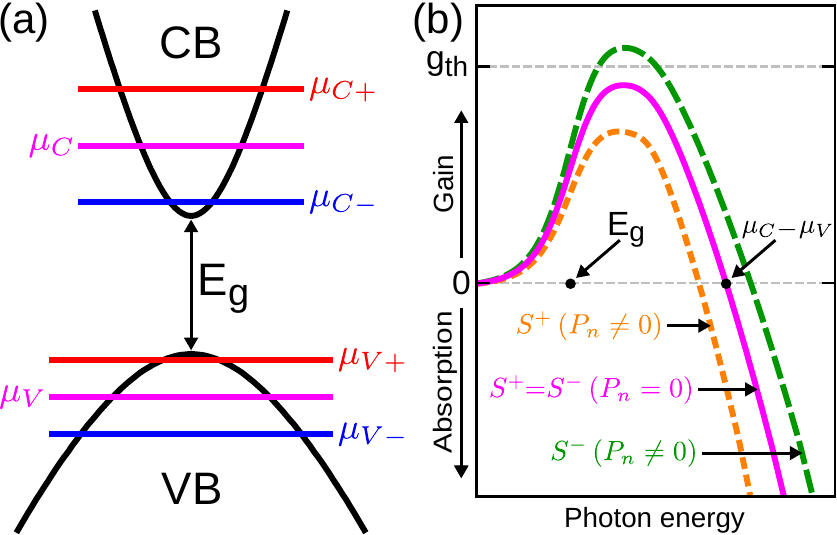}
\caption{(a) Energy band diagram with a bandgap $E_g$ and chemical potentials in 
conduction (valence) bands, $\mu_C$ ($\mu_V$) that in the presence of spin-polarized 
carriers become spin-dependent: $\mu_{C(V)+} \neq \mu_{C(V)-}$, unlike the rest
of our analysis, here holes are spin-polarized. 
(b) Gain spectrum 
for unpolarized (solid) and spin-polarized electrons (dashed curves). 
Positive gain corresponds to the 
emission and negative gain to the absorption of photons. The gain threshold, $g_{th}$, 
for lasing, is attained only for $S^-$ helicity of light. From Ref.~\cite{FariaJunior2015:PRB}.} 
\label{fig:chemical_gain}
\end{center}
\end{figure}

To see how spin-polarized carriers could influence the gain, we show chemical potentials, 
$\mu_{C(V)}$, for a simplified conduction (valence) band in Fig.~\ref{fig:chemical_gain}(a). 
To consider the spin imbalance in the active region, we introduce $\mu_{C\pm}$, 
\begin{equation}
n_\pm= \int_0^\infty dE  \: \rho_{C\pm}(E)/ [1+ \exp(E-\mu_{C\pm})/k_B T],
\label{eq:chem}
\end{equation}
where $E$ is the energy, $\rho_{C\pm}(E)$  is density of states in the conduction band for the respective spin~\cite{FariaJunior:2016}, 
$k_B$ is the Boltzmann 
constant, and $T$ the absolute temperature. Unlike the conventional
chemical potential, $\mu_{C\pm}$ cannot be determined independently because each band does not contain pure
spin states, but mixed states. However, when the spin mixing is not significant, we could assume that  $\mu_{C\pm}$ are
almost independent. Similarly, one can introduce $\mu_{V\pm}$, but we mostly consider that holes are not spin-polarized,
$\mu_{V+}=\mu_{V-}=\mu_V$.

Such different chemical potentials lead to the dependence of gain on the polarization 
of light, described in Fig.~\ref{fig:chemical_gain}(b). Without spin-polarized 
carriers, the gain is the same for $S^+$ and $S^-$ helicity of light. In an ideal 
semiconductor laser, $g>0$ requires a population inversion for photon energies, 
$E_g< \hbar \omega < (\mu_C-\mu_V)$. However, a gain broadening is inherent to 
lasers and, as depicted in Fig.~\ref{fig:chemical_gain}(b), $g>0$ even below the 
bandgap, $\hbar \omega<E_g$. If we assume $P_n\equiv (n_+-n_-)/n\neq0$ 
and $P_p=0$ (accurately satisfied, as spins of holes relax much faster than electrons), 
we see different gain curves for $S^+$ and $S^-$. The crossover from emission to 
absorption is now in the range of ($\mu_{C-} - \mu_{V-}$) and ($\mu_{C+} - \mu_{V+}$).

\begin{figure}[h]
\begin{center}
\includegraphics{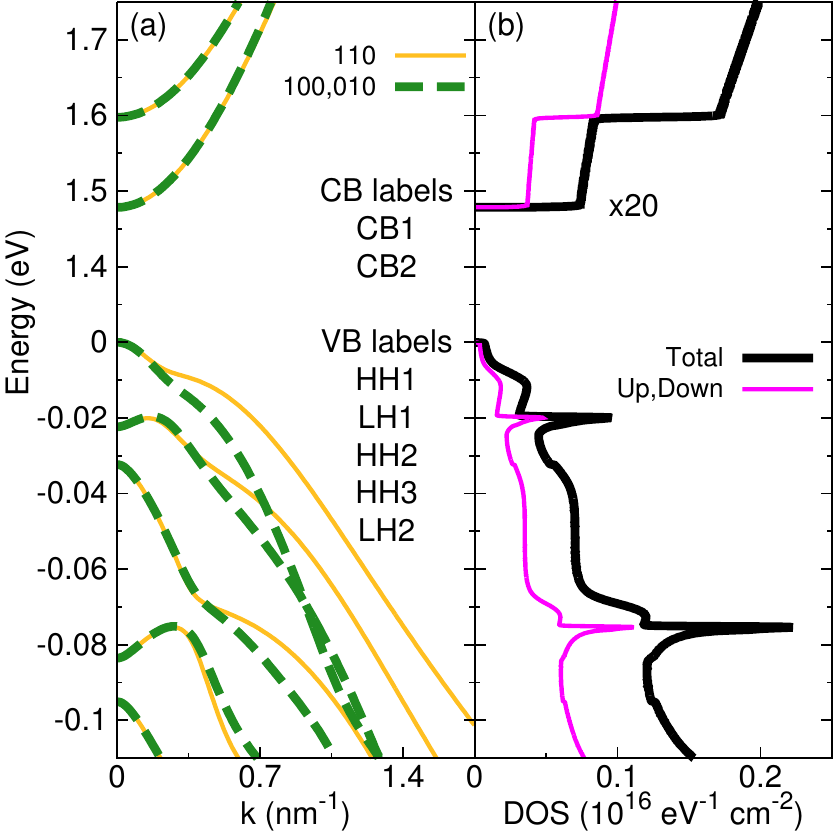}
\caption{(a) Band structure for the $\textrm{Al}_{0.3}\textrm{Ga}_{0.7}\textrm{As}$/GaAs 
QW for different $k$-directions along [100],  [010],  and [110]. 
(b) DOS calculated from (a). 
The conduction band DOS is multiplied by a factor of 20 to match the valence band scale. 
The bandgap is $E_g=1.479\; \textrm{eV}$ 
(CB1-HH1 energy difference).
From Ref.~\cite{FariaJunior2015:PRB}.}
\label{fig:bs_dos}
\end{center}
\end{figure}

For our microscopic description of spin-lasers we focus on a (Al,Ga)As/GaAs-based 
active region, a choice similar to many commercial VCSELs. We consider an $\textrm{Al}_{0.3}\textrm{Ga}_{0.7}\textrm{As}$ 
barrier and a single 8 nm thick GaAs QW. The corresponding 
electronic structure of both band dispersions and the density of states (DOS) is 
shown in Fig.~\ref{fig:bs_dos}. Our calculations, yield two 
confined CB subbands: CB1, CB2, and five VB subbands, labeled in Fig.~\ref{fig:bs_dos}(a) 
by the dominant component of the total envelope function at $\vec{k}=0$, heavy and light holes (HH, LH). 
The larger number of confined VB subbands (with lifted HH/LH degeneracy) 
stems from larger effective masses for holes than 
electrons. These differences in the effective masses also appear in 
the DOS shown in Fig.~\ref{fig:bs_dos}(b). 

To describe the gain spectrum, once we have the 
electronic structure, it is important to understand the effects associated with 
carrier occupancies through injection/pumping. 
The carrier population~\cite{Coldren:2012} is given 
 using the product of the Fermi-Dirac 
distribution and the DOS for CB and VB for both spin projections.

\subsubsection{4c. Spin-Dependent Gain and Birefringence}
\label{bir}

From the conservation of angular momentum and polarization-dependent optical transitions
we can understand that even in conventional lasers 
carrier spin plays a role in determining the gain. 
However, in the absence of spin-polarized carriers (without a magnetic region or an applied magnetic field), 
the gain is identical for the two helicities: $g^+=g^-$, and we recover a simple 
description (spin- and polarization-independent) from Fig.~\ref{fig:gain}(b).  In our notation, 
$g^\pm_\pm$,   the upper (lower) index refers to the circular polarization
(carrier spin) [recall Eq.~(\ref{eq:gain_sum})].

\begin{figure}[h]
\begin{center}
\includegraphics{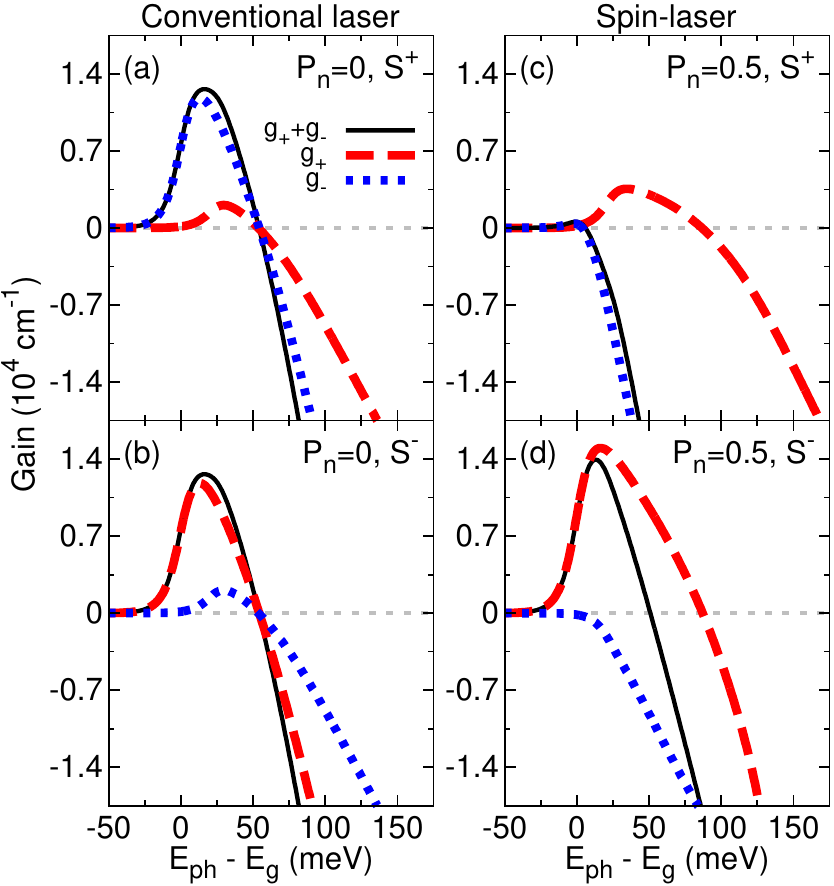}
\caption{Gain spectra shown as a function of photon energy measured 
with respect to the energy bandgap. Conventional laser, $P_n=0$ for (a) $S^+$ 
and (b) $S^-$ light polarization. Spin-lasers, $P_n=0.5$ for (c) $S^+$ and (d) $S^-$ 
light polarizations. For carrier density $n=p=3\times10^{12}\;\textrm{cm}^{-2}$, 
$T=300\;\textrm{K}$, and  $\cosh^{-1}$ broadening with a full width at half-maximum 
of 19.75 meV. From Ref.~\cite{FariaJunior2015:PRB}.}
\label{fig:spin_gain}
\end{center}
\end{figure}

This behavior can be further understood from the gain spectrum in Figs.~\ref{fig:spin_gain}(a) 
and (b), where we recognize that $g^+=g^-$ requires: (i) $g^+_- = g^-_+$ and $g^+_+ = g^-_-$, 
dominated by CB1-HH1 ($1.479\;\textrm{eV} = E_g$) 
and CB1-LH1 ($1.501 \; \textrm{eV}$) 
transitions, respectively (recall Fig.~\ref{fig:bs_dos}).  
No spin-imbalance implies 
spin-independent $\mu_C$ and $\mu_V$ [Fig.~\ref{fig:chemical_gain}(a)] and thus $g^\pm$, $g^\pm_+$, and 
$g^\pm_-$, all vanish at
the photon energy E$_\textrm{ph}=\hbar \omega = \mu_C - \mu_V$. Throughout our calculations 
we choose a suitable $\cosh^{-1}$ broadening~\cite{Chow:1999}, 
which accurately recovers the gain of conventional (Al,Ga)As/GaAs QW systems.

We next turn to the gain spectrum of spin-lasers. Why is their output different 
for $S^+$ and $S^-$ light, as depicted in Fig.~\ref{fig:gain}(b)? Changing only $P_n=0.5$ from 
Figs.~\ref{fig:spin_gain}(a) and (b), we see very different results for $S^+$ and 
$S^-$ light in Figs.~\ref{fig:spin_gain}(c) and (d). $P_n>0$ implies that 
$\mu_{C+}>\mu_{C-}$, leading to a larger recombination 
between the spin up carriers ($n_+p_+ > n_-p_-$) and thus to a larger $g_+$ for 
$S^+$ and $S^-$ (red/dashed line) than $g_-$ (blue/dashed line). The combined 
effect of having spin-polarized carriers and different polarization-dependent optical 
transitions for spin up and down recombination is then responsible for $g^+ \neq g^-$, 
given by solid lines in Figs.~\ref{fig:spin_gain}(c) and \ref{fig:spin_gain}(d). 
For this case, the emitted light $S^-$ exceeds that with the opposite helicity, $S^+$, 
there is a gain asymmetry (also known as the gain anisotropy)~\cite{Holub2007:PRL,Hovel2008:APL,Basu2009:PRL}, 
another consequence of the polarization-dependent gain. 
The gain asymmetry is crucial for spin-selective properties of lasers, including robust spin-filtering 
or spin-amplification, in which even a small $P_n$ (few percent) in the active 
region leads to an almost  complete polarization of the emitted light (of just one 
helicity)~\cite{Iba2011:APL}. Interestingly, with simultaneous polarization of electrons and holes, 
which could be feasible for GaN-based lasers, such gain asymmetry can even change the sign 
by simply changing the total carrier density~\cite{FariaJunior2017:PRB}. 

The zero gain is attained at $\mu_{C+} - \mu_{V}$ for spin up (red curves) and $\mu_{C-} - \mu_{V}$ 
for spin down transitions (blue curves). The total gain, including both of these 
contributions, reaches zero at an intermediate value. Without any changes to the 
band structure, a simple reversal of the carrier spin-polarization, 
 $P_n\rightarrow -P_n$, reverses the role of  preferential light polarization.

An important implication of an anisotropic dielectric function is the phenomenon
of birefringence in which the refractive index, and thus the phase velocity of light, 
depends on the polarization of light~\cite{Coldren:2012}. 
Due to phase anisotropies in the laser cavity and the
polarization-dependence of the refractive index, the emitted frequencies of linearly 
polarized light in the x- and y-directions ($S^x$ and $S^y$) are usually different. 
Such birefringence is often undesired in 
conventional lasers since it is the origin for the typical complex polarization 
dynamics and chaotic polarization switching behavior in 
VCSELs~\cite{SanMiguel1995:PRA,vanExter1997:PRB,Sondermann2004:IEEE,Virte2013:NP,Al-Seyab2011:IEEEPJ}. 
While strong values of birefringence are usually considered to be an obstacle for 
the polarization control in spin-polarized lasers~\cite{Hovel2008:APL,Frougier2015:OE,Yokota2017:IEEEPTL,Fordos2017:PRA},
the combination of a spin-induced gain asymmetry
with birefringence in spin-VCSELs 
allows us to generate fast and controllable oscillations between $S^+$ and $S^-$ 
polarizations~\cite{Gerhardt2011:APL,Hopfner2014:APL}. The frequency of these 
polarization oscillations is determined by the linear birefringence in the VCSEL 
cavity and can be much higher than $\omega_R$.
This opens the path towards ultrahigh bandwidth operation for optical 
communications~\cite{Lee2014:APL,Gerhardt2011:APL,Lindemann2019:N,Gerhardt2012:AOT}, demonstrated in Section~5.

In order to investigate birefringence effects in the active region of a conventional 
laser, we consider uniaxial strain by extending 
the lattice constant  in x-direction. 
For simplicity, we assume the barrier to have the same lattice constant as GaAs, $5.6533 \; \textrm{\AA}$, 
in y-direction. 
For $a_x = 5.6544 \; \textrm{\AA}$ we have the corresponding 
element of the strain tensor $\varepsilon_{xx} \sim 0.019 \%$, while $a_x = 5.6566 \; \textrm{\AA}$ 
gives $\varepsilon_{xx} \sim 0.058 \%$. Besides the differences induced in the band structure, the uniaxial strain also 
induces a change in the dipole selection rules between $S^x$ and $S^y$ light polarizations, 
which can be seen in the gain spectra  $g^x  \neq g^y$.

\begin{figure}[h]
\begin{center}
\includegraphics*[width=8cm]{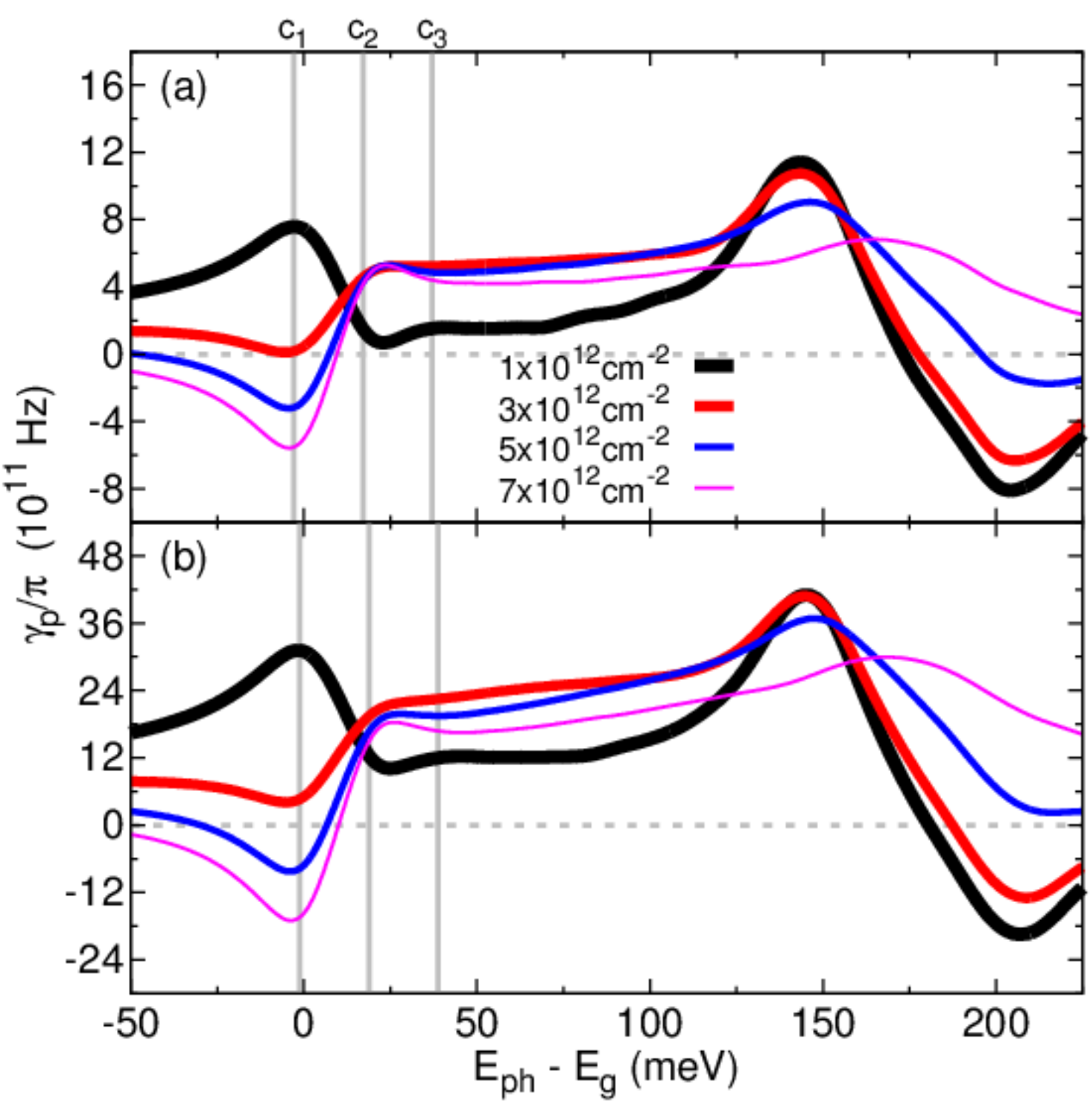}
\caption{Birefringence coefficient as a function of photon energy 
considering (a) $\varepsilon_{xx} \sim 0.019 \%$ and (b) $\varepsilon_{xx} \sim 0.058 \%$. 
Just an increase of $0.0022 \; \textrm{\AA}$ in $a_x$ increases $\gamma_p$ by approximately 
3 times. The two peaks,  around $\textrm{E}_{\textrm{ph}}-\textrm{E}_{\textrm{g}} \sim 0$ 
meV and $\textrm{E}_{\textrm{ph}}-\textrm{E}_{\textrm{g}} \sim 150$ meV are related 
to transitions from CB1 and CB2. Transitions related to CB2 are in the absorption regime.
From Ref.~\cite{FariaJunior2015:PRB}.}
\label{fig:bir_gammap}
\end{center}
\end{figure}

To calculate the birefringence coefficient in the active region, we use the definition~\cite{Mulet2002:IEEEJQE}, 
\begin{equation}
\gamma_{p}(\omega)=-\frac{\omega}{2n_{e}n_{g}}\delta\varepsilon_{r}(\omega) \; ,
\label{eq:gammap}
\end{equation}
where $\omega$ is the frequency of the longitudinal mode in the cavity, 
$n_e$ the effective index of refraction of the cavity and $n_g$ the group refractive 
index. For simplicity, we assume $n_e=n_g$. The real part of the dielectric function 
can be obtained from the imaginary part using the Kramers-Kronig relations~\cite{Chuang:2009}.
We consider three 
resonant cavity positions: $\textrm{c}_1$, $\textrm{c}_2$, 
$\textrm{c}_3$ (vertical lines in Fig.~\ref{fig:bir_gammap}), defining the corresponding energy of emitted photons
$\textrm{c}_1=1.479$ 
eV (CB1-HH1 transition), $\textrm{c}_2=1.501$ eV (CB1-LH1 
transition) and $\textrm{c}_3=1.52\;\textrm{eV}$ (at the high energy side of the 
gain spectrum) and include four carrier densities.

We present the birefringence coefficient in Fig.~\ref{fig:bir_gammap}(a) and \ref{fig:bir_gammap}(b) 
for $\varepsilon_{xx} \sim 0.019 \%$ and $\varepsilon_{xx} \sim 0.058 \%$, respectively. 
This strain in the active region is responsible for modest
changes in bandgap (from 1.483 eV to 1.481 eV) and the gain spectra~\cite{FariaJunior2015:PRB}.
However, it also produces birefringence values of the order of $10^{11-12}$ Hz 
which can be exploited to generate fast polarization oscillations~\cite{Lindemann2019:N}. 
Increasing the strain amount by $\sim 0.04 \%$ from case (a) to case (b), 
the value of $\gamma_p$ increases by approximately 3 times. We also included in 
our calculations spin-polarized electrons and notice that they have only 
a small influence in the birefringence coefficient. Although they slightly change 
$|g^x|$ and $|g^y|$, the asymmetry 
is not affected at all for small $P_n$ of 10-20 \%, relevant values in real devices.
Comparing different cavity designs we observe that for $\textrm{c}_1$, 
the value of $\gamma_p$ strongly decreases and also changes sign with the carrier density, $n$. 
In contrast, for $\textrm{c}_2$ and $\textrm{c}_3$, $\gamma_p$ is always positive. 
For large anisotropies in the DBR,  the birefringence coefficient is on the order of $10^{10}$ Hz, consistent with the 
measurements given in Ref.~\cite{vanExter1997:PRB}. For the 
investigated strain conditions, the main contribution to $\gamma_p$ comes from the 
active region and it is a very versatile parameter that can be fine-tuned using 
both carrier density and cavity designs, possibly even changing its sign and reaching 
carrier density-independent regions~\cite{FariaJunior2015:PRB}.

Even though many-body effects have not been included in the studies of highly-birefringent spin-lasers,
some guidance of their influence is available from 
conventional lasers. For example, we expect band-gap renormalization and Coulomb enhancement 
which typically lead to the red shift and enhancement of the 
gain spectra with the threshold reduction, as compared to the results from the free-career 
model~\cite{Haug:2004,Chow:1999,Burak2000:PRA,Burak2000:IEEEJQE}. However, comparing many-body and free-career models in
conventional lasers reveals only a moderate change in the calculated birefringence~\cite{Burak2000:PRA}.
This suggests that our description of spin-lasers can provide valuable guidance in exploring birefringence, while
in Section~5 we discuss the inclusion of other optical anisotropies.

\subsection{5. Ultrafast Spin-Lasers}

\subsubsection{5a. Birefringent Spin-Lasers}

That the coupling between the spins of carriers and photons could support a high-frequency operation of lasers
was already realized in the first VCSEL with optical spin injection~\cite{Hallstein1997:PRB}. 
The transfer of a coherent Larmor precession of the electron spin to the spin of photons was shown to support polarization
oscillation of the emitted light up to 44 GHz in an external magnetic field of 4 T at 15 K~\cite{Hallstein1997:PRB}. While this approach 
is limited to cryogenic temperatures and does not allow an arbitrary modulation of the polarization, which would be important for 
high-speed communication, nor it is clear if the resulting modulation bandwidth (see Sections~3a and 3b)
could exceed those of conventional lasers, it has motivated other implementations of spin-lasers.

An improved dynamical operation of spin-lasers can be seen from our predictions~\cite{Lee2010:APL} in Section~3b.
With spin injection there is a threshold reduction which enhances the resonance (relaxation-oscillation) frequency  for intensity modulation ({\em IM})
from Eq.~(\ref{eq:om2}), $f_R=\omega_{R, IM}/2\pi$ and, as in conventional lasers~\cite{Chuang:2009},  enhancing the usable frequency range, 
given by the modulation bandwidth (see Fig.~\ref{fig:HO}),
\begin{equation}
f_\mathrm{3dB} \approx \sqrt{1+\sqrt{2}} f_R.
\label{eq:3dB}
\end{equation}

\begin{figure}[h]
\begin{center}
\includegraphics[width=2\linewidth]{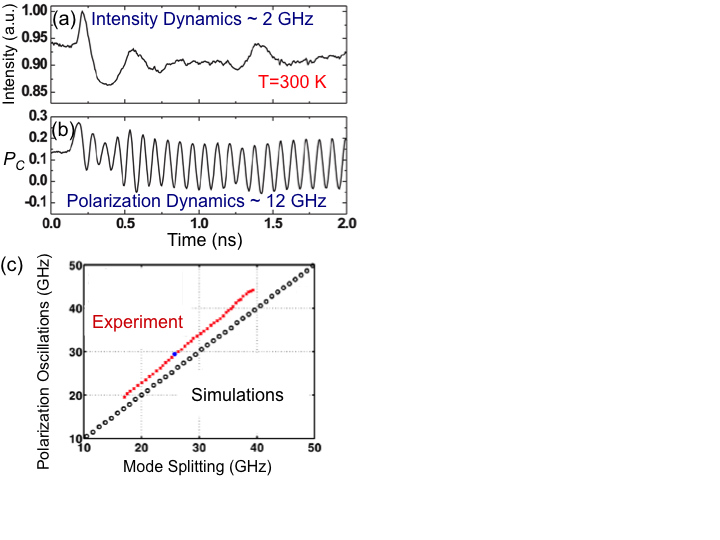}
\vspace{-2cm}
\caption{Oscillations of the emitted light for (a) intensity and (b) photon density polarization ($P_C$)
under hybrid excitation: continuous electrical pumping and pulsed (3 ps) optical pumping with circularly polarized light. 
From Ref.~\cite{Gerhardt2011:APL}. 
(c) A comparison between the birefringence-induced mode splitting and the oscillation frequency of the right circularly polarized 
laser mode. From Ref.~\cite{Lindemann2016:APL}.  
}
\label{fig:bir}	
\end{center}
\end{figure}

In contrast, an enhanced dynamical operation, not relying on a threshold reduction, has been demonstrated by using existing birefringence  
in commercial VCSELs~\cite{Gerhardt2011:APL,Li2010:APL}. A  comparison of Figs.~\ref{fig:bir}(a) and (b) shows that polarization dynamics 
can be much faster than intensity dynamics for a  GaAs-QW VCSEL in which circularly polarized light was used to generate spin-polarized carriers. 
Since the birefringence is responsible for the beating between the emitted light of different helicities,  the changes in the polarization of the emitted light, 
\begin{equation}
P_\textrm{C}=(S^+-S^-)/(S^++S^-),
\label{eq:PC}
\end{equation}
can be faster than the changes in the light intensity.

Even though these experiments have not reached $f_R \sim 20-30$ GHz in the best commercial semiconductor 
lasers~\cite{Haghighi2018:ISLC}, together with theoretical predictions suggesting strain-enhanced much higher operation frequency 
in Section~4c~\cite{FariaJunior2015:PRB}, they provide the proof of concept that birefringence can be useful. Subsequently, 
anisotropic strain was induced into the cavity by pressing a sharp tip close to the gain region~\cite{Lindemann2016:APL}.  The corresponding
results in Fig.~\ref{fig:bir}(c) confirm that oscillation frequency of the $P_C$ grows with the increase of birefringence, 
responsible for the mode splitting, up to 44 GHz, comparable to the fastest $f_R$. 
Even though only supporting static implications of birefringence due to mode splitting, discussed further below, alternative paths
have shown even higher birefringence values using the elasto-optic effect up to $\sim 80$ GHz~\cite{Panajotov2000:APL},  
asymmetric heating up to $\sim 60$ GHz~\cite{Pusch2017:APL},
integrated surface gratings up to 98 GHz~\cite{Pusch2019:EL},  
and mechanical bending reaching 259 GHz~\cite{Pusch2015:EL}. 

\begin{figure}[t]
\centering
\includegraphics*[width=12cm]{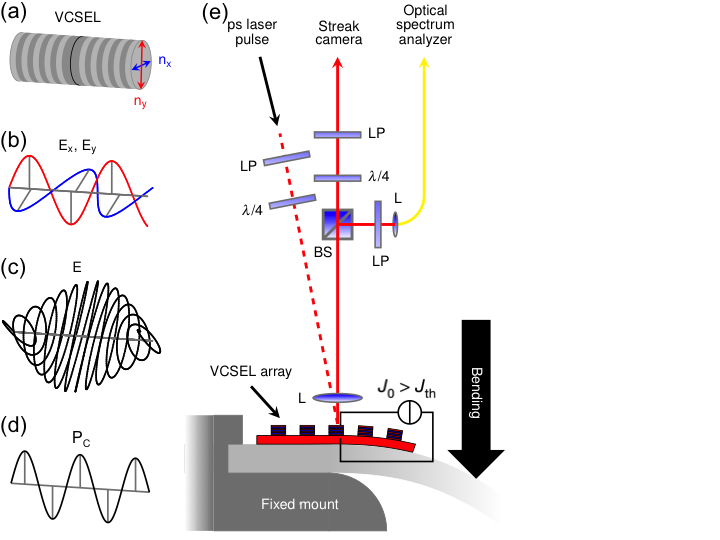}
\caption{Birefringent VCSEL and measurement set-up. (a) VCSEL structure with two polarization-dependent refractive indices, $n_x$ and $n_y$. (b) Linearly polarized electric field ${\bf E}_x$ and ${\bf E}_y$ with frequency difference $\Delta f $ (depicted duration $\ll 1/ {\tilde f}_R$). (c) Total electric field (${\bf E}$) oscillating between right and left circular polarization (depicted duration is $1 /  {\tilde f}_R$). (d) Resulting $P_C$ showing polarization oscillations (depicted duration is $2.5 /  {\tilde f}_R$).  (e) Schematic of experimental design with linear polarizer (LP), quarter-wave plate ($\lambda/4$), lens (L) and beam splitter (BS). The laser is operated with both a pumping current $J_0$ above threshold $J_\mathrm{th}$ and pulsed optical spin injection (dashed red line). The VCSEL emission is depicted as solid red lines and propagates towards both streak camera and fibre (yellow) coupled optical spectrum analyser. The VCSEL sample is bent as illustrated by the black arrow. From Ref.~\cite{Lindemann2019:N}.
}
\label{fig:bent}
\end{figure}

While these experiments on high birefringence are encouraging, it is important to recognize several key  requirements for implementing ultrafast lasers.
To demonstrate that birefringent spin-lasers are indeed suitable for ultrafast polarization modulation and data transmission, four major challenges 
had to be overcome: (1) implementing a large birefringence splitting; (2) proving that birefringence can control the polarization oscillation 
frequency; (3) demonstrating that polarization dynamics in spin-lasers can be as fast as hundreds of GHz, faster than other state-of-the-art
concepts for direct laser modulation; and (4) verifying its capability for optical data communication, including a large modulation bandwidth. 
For a practical implementation it is also important to establish a low-power consumption.

Remarkably, with a highly-birefringent AlGaAs-based 850 nm VCSEL, depicted in Fig.~\ref{fig:bent}, all these challenges can be addressed,
unlike the common expectations that birefringence is detrimental in conventional and spin-lasers~\cite{Hovel2008:APL,Frougier2015:OE,Yokota2017:IEEEPTL,Fordos2017:PRA}.
We  demonstrate room-temperature polarization oscillations (PO) with ${\tilde f}_R > 200$ GHz, 
resulting in a polarization modulation ({\em PM}) bandwidth $> 240$ GHz, an order of magnitude larger values 
than in the best conventional lasers. This operation is achieved using a hybrid pumping scheme 
of a constant electrical injection of spin-unpolarized carriers $J_0$ above the threshold injection $J_{th}$ and 
a ps laser pulse serving as spin injection. The laser pulse becomes circularly polarized after passing through the 
linear polarizer and the quarter-wave plate, which excites spin-polarized carriers in the gain region. 

The cavity anisotropy of the considered VCSELs is responsible for the linear birefringence, $\gamma_p$ (recall Section~4c), 
manifested in (i) linearly polarized emission due to anisotropy of
the refractive index $n_x\neq n_y$, sketched in Fig.~\ref{fig:bent}(a) and (ii) the frequency of mode splitting,   
\begin{equation}
\Delta f \approx \frac{\gamma_p}{\pi}-\frac{\alpha}{\pi} \gamma_a,
\label{eq:delta_f}
\end{equation} 
of the two orthogonal modes, described by electric fields  {\bf E}$_x$, {\bf E}$_y$ in Fig.~\ref{fig:bent}(b), where  
$\alpha$ is the linewidth enhancement factor, and $\gamma_a$ is the dichroism, 
which represents the anisotropy of absorption (or, equivalently, of optical gain). 
A usually weak coupling between the two modes results in an unstable behavior with polarization switching and ${\tilde f}_R$ 
related to the beat frequency between the {\bf E}$_x$ and {\bf E}$_y$ and leads to the periodic evolution of the total electric field, {\bf E}={\bf E}$_x$+{\bf E}$_y$, 
shown in Fig.~\ref{fig:bent}(c). The resulting $P_\textrm{C}$ can be controlled by {\em PM}, discussed in Eq.~(\ref{eq:PM}).

\begin{figure*}[t]
\centering
\includegraphics*[width=0.82\linewidth]{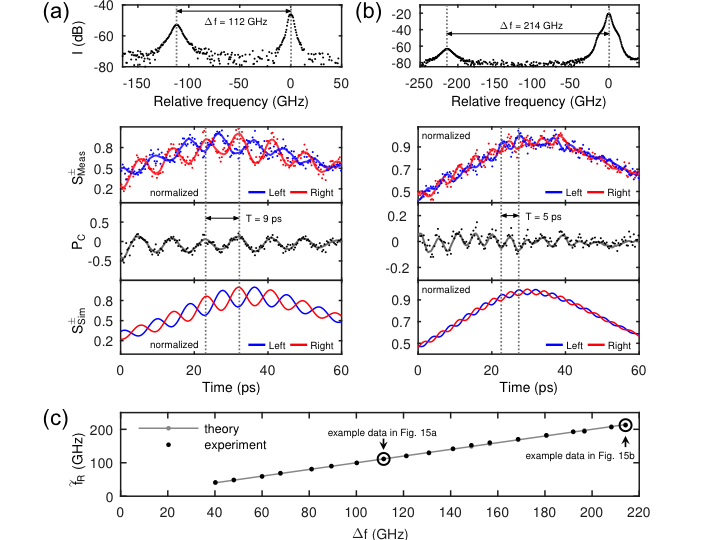}
\caption{Polarization behavior of spin-VCSELs. (a), (b) Top panel, the
optical spectrum I reveals the birefringence-induced mode splitting ($\Delta f$) to be 112 GHz (a) and 214 GHz (b). The main mode
 (at 0 GHz) was suppressed for display. Bottom panel: $S^{\pm}_\mathrm{Meas}$, $P_C$, and $S^{\pm}_\mathrm{Sim}$  are plotted against $S^{\pm}_\mathrm{Meas}$ gives the measured polarization-resolved normalized photon densities after pulsed spin injection (dots, raw data; lines, smoothed data; `Right,' $S^+_\mathrm{Meas,Sim}$ `Left,' $S^-_\mathrm{Meas,Sim}$). $T$ denotes the period of the PO
frequency. From $S^{\pm}_\mathrm{Meas}$, $P_C$ is determined, showing the PO. $S^{\pm}_\mathrm{Sim}$ gives the  simulated behavior. (c) Plot of $f_R$ against $\Delta f$, showing experimental data (dots) and results from theory (grey line). The polarization dynamics can be tuned by birefringence-induced mode splitting. Data in (a) and (b) are part of the tuning series shown in (c) [ringed dots]. At $\Delta f = 214$ GHz, the frequency tuning is stopped to prevent mechanical sample damage. From Ref.~\cite{Lindemann2019:N}.
}
\label{fig:polar}
\end{figure*}

By generalizing a  model for conventional VCSELs and the rate equations (\ref{eq:REn})$-$(\ref{eq:RES}) to include the effects of optical anisotropies, 
as discussed in Section~5b, our theoretical analysis reveals that the PO has a resonant behavior similar to the intensity oscillation in conventional 
VCSELs, but with a different frequency,
\begin{equation}
\tilde{f_R}=\frac{\gamma_p}{\pi}-\frac{\gamma_r S_0}{4\pi(\gamma_s^2+4\gamma_p^2)\tau_p}(\alpha\gamma_s-2\gamma_p)-\frac{\epsilon_p S_0}{4\pi},
\label{eq:fR_tilde}
\end{equation} 
where $\gamma_r$  and $\gamma_s=1/\tau_{sn}$ are the carrier recombination and spin-flip rates, $\alpha$ the linewidth enhancement factor, $\epsilon_p$ is the  phase-related saturation and $S_0$, as in Section~3b,  is the steady-state photon density normalized to its value at 2$J_{th}$. For a large $\gamma_p$ 
Eq.~(\ref{eq:fR_tilde}) is valid for all practical pumping regimes, while the last two terms are negligible compared with $\gamma_p/\pi$. 
Thus a  strongly enhanced birefringence,  $\tilde{f_R}\approx \gamma_p/\pi  \gg f_R$, may overcome the frequency limitations of conventional lasers.

Our results from the measurement setup in Fig.~\ref{fig:bent}(b) are  shown for $\Delta f=112$ GHz  ($\Delta f= 214$ GHz, approaching damage threshold) in 
Figs.~\ref{fig:polar}(a)  and \ref{fig:polar}(b). In the polarization-resolved measured and simulated normalized intensities $S^{\pm}_\mathrm{Meas}$ and $S^{\pm}_\mathrm{Sim}$, the slow envelope is the second peak of the intensity relaxation oscillation after excitation. Its frequency is $f_R\approx 8$ GHz for both datasets. In $P_\mathrm{C}$, the overlaying intensity dynamics vanishes and only the fast PO is evident, showing $\tilde{f}_R= 112$ GHz $\approx 14 f_R$ or $\tilde{f}_R= 212$ GHz $\approx 27 f_R$. The polarization dynamics is more than an order of magnitude faster than the intensity dynamics in the same device! 

The PO amplitude is decreasing with increasing frequency, both due to the bandwidth of the measurement system and due to fundamental limitations of the polarization dynamics including dichroism, spin-flip rate, and pumping current. By changing $\Delta f$ with bending, as depicted in Fig.~\ref{fig:bent}(d), $\tilde{f_R}$ is continuously tuned up to 212 GHz, showing an excellent agreement with the theoretical calculation over the entire 
frequency range in Fig.~\ref{fig:polar}(c). 

For a regime in which the optical anisotropies are dominated by $\gamma_p$, we can approximate Eqs.~(\ref{eq:delta_f}) and (\ref{eq:fR_tilde}) as,
$\Delta f \approx {\tilde f}_R \approx \gamma_p/\pi$. While numerically $\Delta f$  and ${\tilde f}_R$ are closely related, they refer to static ($\Delta f$) and 
dynamic regime (${\tilde f}_R$). Therefore, there is a crucial difference between the static response of birefringence, visible as mode 
splitting in the spectrum of the laser emission reported in Ref.~\cite{Pusch2015:EL}, and the possibility of achieving high-frequency operation of a spin-VCSEL 
discussed here. 

\subsubsection{5b. Generalized Spin-Flip Model}

To study the dynamics of highly-birefringent spin-lasers it is important to consider
optical anisotropies missing in Eqs.~(\ref{eq:REn})$-$(\ref{eq:RES}). A suitable starting
point is  the spin-flip model~\cite{SanMiguel1995:PRA,Al-Seyab2011:IEEEPJ,Alharthi2015:APL}, initially used to study dynamical 
operation and linear polarization switching in conventional VCSELs~\cite{SanMiguel1995:PRA}. 
The framework for this model comes from the Maxwell-Bloch 
equations~\cite{Haug:2004,Chow:1999,SanMiguel1995:PRA,Fordos2017:PRA,Mulet2002:IEEEJQE,Burak2000:PRA,Burak2000:IEEEJQE, Hess1996:PRA,Bowden1993:OC}
which describe interaction between the gain medium and electromagnetic field within the dipole approximation for optical transitions by relating electric field, 
polarization, and carrier densities. However, for spin-lasers various generalizations
of the spin-flip model~\cite{SanMiguel1995:PRA} are needed, such as including unequal spin-relaxation rates for electrons and holes, as in Section~2c.
Unlike the photon densities (intensities)  $S^\pm$ used in  (\ref{eq:REn})$-$(\ref{eq:RES}), the spin-flip model
is described with normalized helicity amplitudes of the electric field, $E^\pm$, where $S^\pm =|E^\pm|^2$. The resulting equations are~\cite{Lindemann2019:N},
\begin{eqnarray}
\label{eq:SFME}
\dot E^{\pm}&=&(1/(2\tau_{ph}))  (1+i\alpha) (N\pm n-1)E^{\pm}-(\gamma_a+i \gamma_p ) E^{\mp}   \nonumber \\
&-& (\epsilon_a + i \epsilon_p) \vert E^{\pm} \vert^2 E^{\pm},  \\
\label{eq:SFMN}
\dot N&=&\gamma_r\left[J_- (t) +J_+ (t)\right]-\gamma_r N -\gamma_r(N+n)\vert E^{+}\vert^2 \nonumber \\
&-&\gamma_r(N-n)\vert E^{-}\vert^2, \\
\label{eq:SFMn}
\dot n&=&\gamma_r\left[J_- (t) -J_+ (t)\right]-\gamma_s n-\gamma_r(N+n)\vert E^{+}\vert^2  \nonumber \\
&+&\gamma_r(N-n)\vert E^{-}\vert^2, 
\end{eqnarray}
where $E^{\pm}$ are  slowly varying amplitudes of the electric field, 
$\tau_p$ and $J_{\pm} (t)$ retain the meaning from Eqs.~(\ref{eq:REn})$-$(\ref{eq:RES}), while 
$\gamma_r$, $\gamma_p$, $\gamma_a$,  $\gamma_s$, and $\alpha$ were introduced in Eqs.~(\ref{eq:delta_f}) and ~(\ref{eq:fR_tilde}).
$N$ is the total population inversion of laser upper and lower levels, $n$ is the population difference between spin-down and spin-up electrons with a decay rate 
$\gamma_s$, see Eq.~(\ref{eq:REn}). The relation between the $P_\textrm{C}$ and the spin polarization of recombining carriers is given by the standard dipole selection rules in semiconductors. Since holes typically have a negligibly short $\tau_{sp}$, only the electrons are considered spin-polarized, see Section~2c.

In the standard spin-flip model~\cite{SanMiguel1995:PRA}, the polarization of the active region material was assumed to respond linearly to {\bf E}, but that fails to describe our experimental observations.  Therefore we consider a generalized nonlinear response such that the saturation effects are included and polarization $P_{\pm}=(\chi_{\pm} - \tilde \epsilon \vert E^{\pm} \vert^2) E^{\pm} $. Here, an intensity-dependent part with a complex coefficient 
$\tilde \epsilon=\epsilon_a+i\epsilon_p$ is subtracted from the linear susceptibility $\chi_{\pm}$ to quantify saturation effects associated with amplitude and phase of the field, respectively. This leads to the additional term $- (\epsilon_a + i \epsilon_p) \vert E^{\pm} \vert^2 E^{\pm}$ in Eq.~(\ref{eq:SFME}) in comparison to the model used in Ref.~\cite{SanMiguel1995:PRA}.

For our dynamic simulations using the generalized spin-flip model,
the values for important parameters, such as the gain anisotropy, have been
extracted from a series of experimental measurements~\cite{Lindemann2019:N}.
Thus, direct influence of the Coulomb interaction on the steady-state
parameters are effectively considered.

\subsubsection{5c. Ultrafast Modulation Response}

\begin{figure}[h]
\centering
\includegraphics*[width=6.85cm]{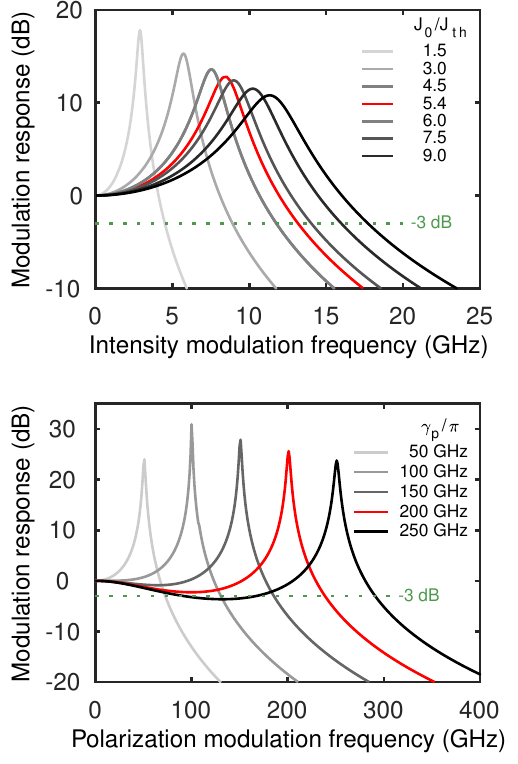}
\vspace{-0.1cm}
\caption{Advantages of polarization modulation ({\em PM}) in dynamic performance.
(a) Simulated intensity modulation ({\em IM}),
$J(t) = J_0 + \delta J \sin(2\pi f t)$, response for varying normalized electric pumping $J_0/J_{th}$, where $J_0$ is the fixed bias current and $\delta J$ is the {\em IM} amplitude. (b) {\em PM}, $P_J(t) = P_0 + \delta P \sin(2\pi f t)$, response for various birefringence conditions (see key). Red traces: the simulations for the 
studied VCSEL at $\gamma_p/\pi$ = 200 GHz. From Ref.~\cite{Lindemann2019:N}.
}
\label{fig:advantage}
\end{figure}

The demonstrated record-high PO frequency in Fig.~\ref{fig:polar} provides the basis to evaluate the performance of spin-lasers for digital optical communication, 
characterized by the modulation bandwidth (see Section~3b) and data transfer rates, commonly analyzed using so-called eye 
diagrams~\cite{Wasner2015:APL,Lindemann2019:N}. 
Modulation bandwidth and data transfer rates are modeled using the methods from Section~5b with a parameter set carefully 
obtained from the experiments.
The IM, given in Fig.~\ref{fig:advantage}(a) by the response function $R(f)=\delta S/\delta J$, similar to Section~3b,  resembles the displacement of a harmonic oscillator from Fig.~\ref{fig:HO}.  The bandwidth given by $f_\mathrm{3dB}$ in Eq.~(\ref{eq:3dB}) is enhanced, just as $f_R$, by the increasing photon density through increased pumping current. 
In our experiments, the pumping current of 5.4$J_{th}$ corresponds to $f_\mathrm{3dB}\approx13.5$ GHz, while $f_{3dB}<20$ GHz for all 
pumping rates $J_0/J_{th}$. 
Remarkably, for the same device parameters, Fig.~\ref{fig:advantage}(b) reveals 

\noindent{a} huge increase in the {\em PM} bandwidth from the response function 
$R(f)=\delta P_\textrm{C}/\delta P_J$. Similar to $\tilde{f_R}$ in Eq.~(\ref{eq:fR_tilde}), the corresponding bandwidth $\tilde{f}_\mathrm{3dB}$ increases with $\gamma_p$.

\begin{figure}[h]
\centering
\includegraphics*[width=12cm]{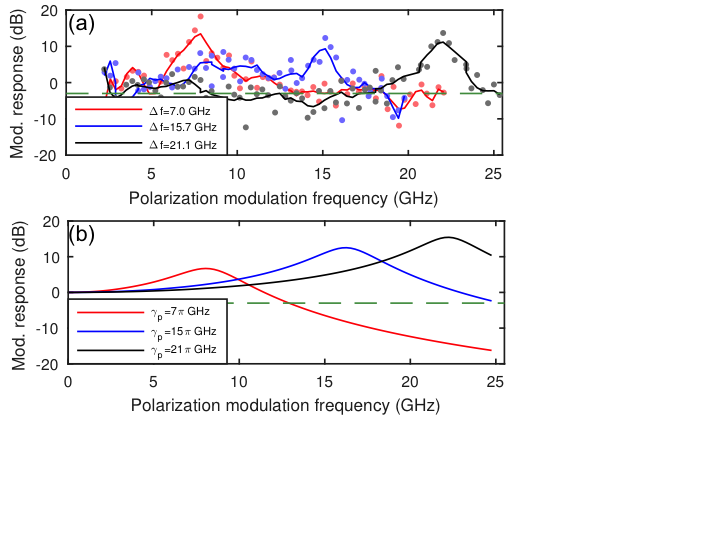}
\vspace{-2.4cm}
\caption{Experimental {\em PM} and its birefringence dependence. The optical pulsed spin injection from Fig.~\ref{fig:bent} 
is replaced by a diode laser in continuous wave operation. 
(a) Results for several values of the birefringence-induced mode splitting, $\Delta f$. Points mark raw data, lines are smoothed data. 
(b) Numerical verification of results in (a) using the parameter set obtained in this work for the appropriate values of $\gamma_p$.  
From Ref.~\cite{Lindemann2019:N}.
}
\label{fig:PM}
\end{figure}

While readily available excitation pulses can 
trigger the PO observed in Fig.~\ref{fig:polar}, experimental demonstration of a huge increase in the {\em PM}  bandwidth from Fig.~\ref{fig:advantage} is much more challenging. 
For that purpose a spin-VCSEL is modulated using a semiconductor laser diode with a continuous wave operation and an additional {\em PM} 
setup containing an electrooptic modulator. The maximum modulation is limited by the modulator to $\sim25$ GHz, the fastest available commercial device at the considered frequency for our AlGaAs lasers.
With this set-up $P_\textrm{C}$ is modulated, while the intensity stays constant.

The corresponding experimental data in Fig.~\ref{fig:PM}(a) which
contain response resonance peaks at frequencies corresponding to the birefringence-induced mode splitting. This proves that the resonance frequency of the polarization modulation response can be tuned by the birefringence-induced mode splitting as well as the POs. The simulations in 
Fig.~\ref{fig:PM}(b) using parameter set obtained in this work and methods from Section~5b resemble the resonance peaks from the experiment. 

\begin{figure}[h]
\centering
\includegraphics*[width=11cm]{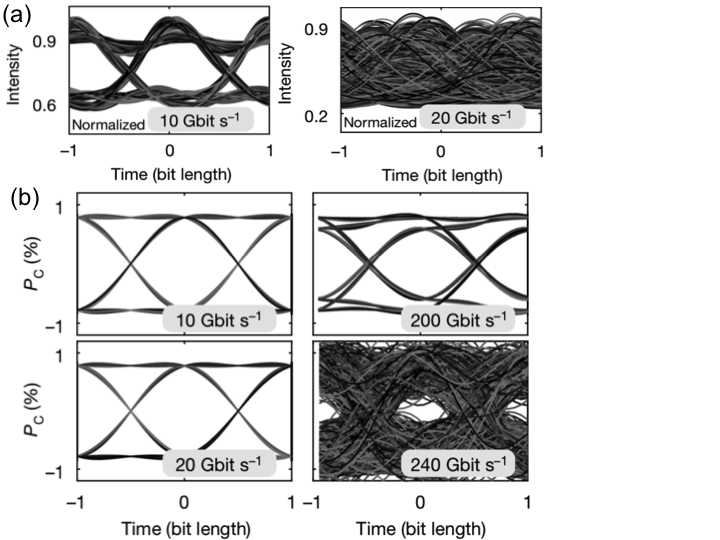}
\vspace{-0.5cm}
\caption{Advantages of {\em PM} in eye diagrams.
(a) {\em IM} eye diagrams  
using filtered pseudorandom bit sequences for 10 Gbit s$^{-1}$ and 20 Gbit s$^{-1}$. 
The intensity plots are normalized to their maximum values. (b) {\em PM} 
eye diagrams showing circular polarization 
for bit sequences (in Gbit s$^{-1}$) of 10, 20, 200 and 240 in the same device at 
$\gamma_p/\pi$ = 200 GHz. From Ref.~\cite{Lindemann2019:N}.
}
\label{fig:eye}
\end{figure}

\begin{figure}[h]
\centering
\includegraphics*[width=15.5cm]{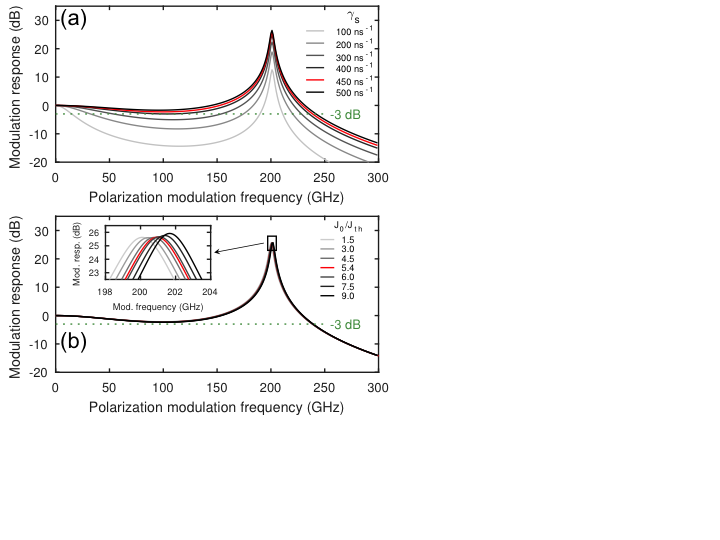}
\vspace{-3cm}
\caption{Influences on modulation bandwidth. (a) Influence of spin-flip rate $\gamma_s$ on modulation bandwidth, shown by plotting modulation response as a function of polarization modulation frequency. (b) As (a) but showing the influence of normalized electric bias 
$J_0/J_{th}$. Inset, magnified view of boxed area in main plot, showing a nearly pumping-independent resonance frequency. Red traces mark the simulations for the studied VCSEL at 
$\gamma_p/\pi = 200$ GHz. From Ref~\cite{Lindemann2019:N}.
}
\label{fig:bandwidth}
\end{figure}

The advantages of highly-birefringent spin-VCSELs extend also to digital operation with abrupt 
changes between {\em off} and {\em on} injection (recall Section~3c). For {\em IM},
\begin{equation}
J(t)=J_0+\delta J(t)=J_0+\delta J f(t),
\label{eq:jmod}
\end{equation}
where $\delta J$ is the  modulation amplitude and $f(t)$ a binary function 
of the coded input signal, while analogous expression could be written for {\em PM}.
The corresponding  intensity or circular polarization of emitted light reflects information encoded into the input injection.
For {\em IM}, bits `0' and `1' are defined by being above or below a specified threshold in the light intensity.
For {\em PM}, `0'  and `1' are encoded as left and right circular polarization. 

To analyze the quality of digital data transfer, it is convenient to use eye diagrams, widely employed
in conventional lasers~\cite{Michalzik:2013} and introduced in spin-lasers~\cite{Wasner2015:APL},
where a binary signal, $f(t)$ in Eq.~(\ref{eq:jmod}), is simulated by 2$^{10}$ pseudorandom bits. 
An eye diagram is generated by dividing the laser wave form into
segments of an equal bit size (two bits in Fig.~\ref{fig:eye}) and overlaying them. 
A bit slot time (bit length) is an inverse of a bit rate. 
The central opening of  an eye diagram resembles a human eye; a larger opening
signifies a higher quality of a digital signal in terms of the reduced errors and noise.
The width of an opening defines the time over which the received signal can be sampled
error-free from intersymbol interference, while its height defines the noise margin.

For the same set of VCSEL parameters and $\gamma_p/\pi = 200$ GHz, {\em IM} and {\em PM} are compared 
in Figs.~\ref{fig:eye}(a) and (b).  The closing eye diagram precludes data transfer by {\em IM}  slightly above 10 
Gbit s$^{-1}$. In contrast, looking at the eye diagrams, we see that {\em PM} supports data transfer up to 240 Gbit s$^{-1}$, 
showing a remarkable improvement in digital operation over conventional VCSELs, consistent with the increase of
${\tilde f}_\mathrm{3dB}$ over $f_\mathrm{3dB}$ from Fig.~\ref{fig:advantage}.

The principle of improving spin-lasers by an enhanced birefringence~\cite{Gerhardt2011:APL,FariaJunior2015:PRB,Lindemann2019:N} can be extended to other devices
and wavelengths.  
Recent experimental results by Yokota and collaborators in InAlGaAs QW-based VCSELs~\cite{Yokota2018:CLEO,Yokota2018:APL}
verify that the {\em PM} bandwidth  is enhanced by birefringence. With birefringence-induced splitting of 19 GHz the {\em PM} bandwidth reaches 
23 GHz~\cite{Yokota2018:CLEO}. This demonstration is important as it is realized for 1.55 $\mu$m emission, the optimal wavelength for long-haul communications, 
confirming the importance of tailoring polarization dynamics and spin imbalance in lasers. 

Unlike common approaches in spintronics and spin-lasers~\cite{Zutic2004:RMP,Iba2011:APL} that seek to increase the spin relaxation time ($1/\gamma_s$), 
we find that short spin relaxation times are desirable for {\em PM}.  With sufficiently fast spin relaxation, as long as $\gamma_s$ is equal or larger than $2\gamma_p/\pi$, Fig.~\ref{fig:bandwidth}(a) reveals that the modulation response at $f<\tilde{f_R}$ remains above $-3$ dB, and
therefore the {\em PM} bandwidth exceeds the $\tilde{f_R}$. While in our VCSEL the spin-flip rate is close to its optimum value for a $\tilde{f_R}$ slightly above 200 GHz,
this dependence of {\em PM} on $\gamma_s$ confirms the importance of addressing multiple challenges for ultrafast spin-lasers, noted in Section~5a. Just high-$\gamma_p$ values
are not sufficient, even if they lead to high-frequency PO and it is also crucial to consider VCSEL design and other materials parameters for optimal performance. For future spin-VCSELs
seeking to reach THz modulation bandwidth it is encouraging that high $\gamma_s= 1000$ ns$^{-1}$ was measured at 300 K for GaAs VCSELs~\cite{Blansett2001:OE}, 
allowing for $\tilde{f_R}> 500$ GHz, while even higher $\gamma_s$ is possible in (In,Ga)As devices~\cite{Kini2008:JAP}.

Another peculiar {\em PM} trend, as shown in Fig.~\ref{fig:bandwidth}(b), 
comes from the corresponding bandwidth dependence on the pumping power.  
The push for faster conventional VCSELs as well as other photonic devices typically requires a stronger pumping for {\em IM}, which leads to fundamental limitations. 
For example, simulating the device while neglecting heating effects, an increase in pumping from $1.5 J_{th}$ to $9 J_{th}$ enhances $f_R$ from 2.8 to 11.3 GHz and $f_{3dB}$ 
from 4.5 to 17.8 GHz. However, higher pumping generates higher dissipated power which increases the laser temperature. In contrast, for {\em PM} in 
Fig.~\ref{fig:bandwidth}(b)  $\tilde{f}_R$ is  almost independent of pumping. Thus, the highest bit rates can be already attained slightly above threshold. 
This has a great potential for ultra-low-power optical communication. In a conventional 850 nm VCSEL, a heat-to-data ratio $\rm{HDR}=56$ fJ/bit 
at 25 Gbit/s was demonstrated~\cite{Moser2012:EL}. Utilizing PM for our devices, assuming pumping at $1.5 J_{th}$ with electrical spin-injection, 
a much lower $\rm{HDR}=3.8$ fJ/bit could be obtained at a substantially higher bit rate of 240 Gbit/s.

\subsection{6. Conclusions and Outlook}

\begin{figure*}[t]
\begin{center}
\includegraphics[width=0.8\linewidth]{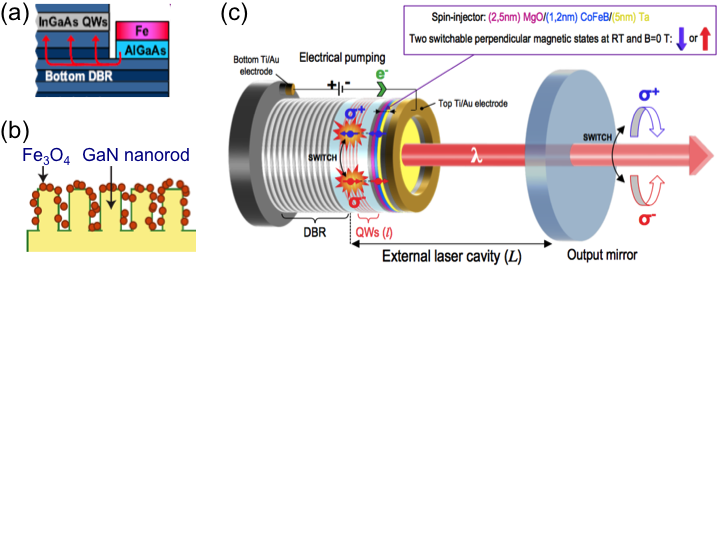}
\vspace{-5.5cm}
\caption{(a) Electrical spin injection in VCSEL from Fe spin injector to the gain region, InGaAs QWs.
From  Ref.~\cite{Holub2007:PRL}.
(b) Integrating Fe$_3$O$_4$ nanomagnets with the gain region, GaN nanorods. From Ref.~\cite{Cheng2014:NN}.
(c) A proposal for electrically injected spin-laser with an external cavity. With a perpendicular magnetization, 
 in the absence of applied magnetic field it would be possible to switch between injected spin-up/down 
 electrons and emit light of different helicity. From Ref.~\cite{Frougier2015:OE}.}
\label{fig:outlook}
\end{center}
\end{figure*}

While the study of spin-lasers is currently conducted by only a modest number of research groups, 
it offers fascinating opportunities building on advances in spintronics and conventional lasers. 
As in the research on their conventional counterparts, one can envision that spin-lasers could provide 
a versatile platform to study fundamental phenomena, from phase transitions to chaos~\cite{Degiorgio1970:PRA,Erneaux:2010}.
These possibilities are now even richer by the control of spin degrees of freedom~\cite{Raddo2017:SR}. Furthermore, the instabilities 
found in lasers directly resemble instabilities in electronic devices~\cite{Degiorgio1976:PT}.

Since lasers provide highly-accurate and tunable parameters, important insights can be achieved by establishing mapping procedures 
between lasers and other cooperative phenomena, such as ferromagnetism~\cite{Haken:1985,Degiorgio1970:PRA, Degiorgio1976:PT}. 
The concept of lasing is not limited to photons and there is a growing interest in the lasing of polaritons, quasiparticles resulting from 
the coupling between excitons and confined photons~\cite{Deng2010:RMP}. The control of microcavity polaritons is considered for 
exploring quantum phase transitions and realizing Bose-Einstein condensation in solids~\cite{Deng2010:RMP,Schneider2013:N}.
The demonstration of room-temperature spin-polariton lasers~\cite{Bhattacharya2017:PRL}, reveals novel opportunities to explore 
the implications of spin-polarized carriers and the helicity of light in lasers.

The potential of spin-lasers goes beyond fundamental phenomena, and there are many 
applications where they could offer 
important breakthroughs. A striking example is the growth in communication; from massive data centers, multicore microprocessors, 
and high-performance computing. The resulting limitations in communication and interconnects are the bottleneck in Moore's law scaling 
and the main source of power dissipation~\cite{Hecht2016:N,Miller2017:JLT,Miller2009:PIEEE,Hilbert2011:S,Jones2018:N}. 
Impressive accomplishments in spin-lasers have already been realized. This pertains not only to demonstrating 
performance of lasers superior to their conventional counterparts, as discussed in Section~5, but also to elucidating 
novel concepts and operation principles in spintronics. Simultaneous spin polarization
of electrons and holes, a coupling between spin, carrier, and photon dynamics, amplification,
and strong nonlinear response, offer many unexplored opportunities. The modulation of spin polarization 
in lasers has motivated introducing the concept of spin interconnects~\cite{Dery2011:APL,Zutic2011:NM}
while using the spin-polarized carriers can also enable coherent emission of phonons~\cite{Khaetskii2013:PRL}.   

For practical applications it would be important to go beyond the commonly employed optically-pumped
lasers and realize their room-temperature electrical operation. Some of the related challenges
can be seen from an early demonstration of electrical spin injection in a spin VCSEL~\cite{Holub2007:PRL}, 
depicted in Fig.~\ref{fig:outlook}(a) and also in Fig.~\ref{fig:exp}. With the distance of several $\mu$m between the spin 
injector (Fe) and the gain region [(In,Ga)As QW], through spin relaxation the  spin polarization of injected carriers is substantially 
reduced when it reaches the gain region~\cite{Soldat2011:APL}, limiting the electrical operation to $\sim$100 K~\cite{Holub2007:PRL}. 
Replacing the QW by QD gain region can enhance the operation temperature up to $\sim$230 K using MnAs 
injector~\cite{Saha2010:PRB}.  A strong confinement in QDs can suppress the effect of spin-orbit coupling, a leading mechanism for 
the spin relaxation of carriers and thus enhance the spin relaxation time~\cite{Zutic2004:RMP}. 
While a long spin-relaxation time is needed for a material between the spin injector and the gain region, such that an appreciable 
spin polarization reaches the gain region, we see from Section~5 that in the gain region itself a short spin relaxation time
could be desirable to enhance the modulation bandwidth for highly-birefringent lasers. Experimental room-temperature 
spin injection in MnAs-based spin-LEDs~\cite{Fraser2010:APL} is also encouraging for a room-temperature spin injection in lasers. 

To implement a robust room-temperature electrical spin injection in lasers it will help to use the knowledge from 
spin injection in LEDs, as well as the suitable choice of materials and interface quality between the ferromagnets and semiconductors,
which both determine the efficiency of spin injection~\cite{Hanbicki2002:APL,Hanbicki2003:APL,Zega2005:PRL,Salis2005:APL}. 
Adopting advances in commercial devices based on MgO-barriers for large magnetoresistive effects will be useful~\cite{Tsymbal:2019}. With the spatial 
separation between the spin injector and gain region, it is important to establish an accurate modeling of spin transport. Spintronic devices 
usually rely on magnetoresistive effects and unipolar transport; where only one type of carriers (electrons) plays an active role such 
that the equivalent resistor picture is usually sufficient to describe spin transport and spin injection. 

In contrast, for modeling bipolar spintronic devices~\cite{Zutic2006:IBM,Zutic2007:JPCM}, including 
spin-lasers, it is important to recognize limitations of the equivalent resistor picture~\cite{Zutic2004:RMP}. Deviations from local charge neutrality, 
band bending, nonlinear current-voltage characteristics, an explicit bias-dependence of spin injection and the presence of both 
electrons and holes are neglected in this approach. From the self-consistent description using generalized drift-diffusion 
and Poisson equations~\cite{Zutic2001:PRB,Zutic2006:PRL,Zutic2002:PRL}, one can see that in semiconductor junctions current-voltage 
nonlinearities  are important even at small bias. The spin injection strongly depends on the applied bias~\cite{Crooker2009:PRB,Li2009:APL}, 
not just on the relative resistances of  the two regions.

While almost all spin-lasers have been based on zinc-blende semiconductors, such as GaAs or InAs, 
a spin-laser with a gain region made of a wurtzite semiconductor (GaN-based) has been the 
first case of an electrically manipulated spin-laser at room temperature~\cite{Cheng2014:NN}. Both the polarization
and the intensity of the emitted light were tunable by external bias.
That operation was realized through spin-filtering by integrating nanomagnets with the active 
region of an optically pumped nanorod laser as illustrated in Fig.~\ref{fig:outlook}(b). The nanomagnets were too small
to support ferromagnetism and a modest applied field (0.35 T) was needed to magnetically align them.
Unlike  typical gain regions in which hole spin relaxation times is much shorter than for electrons, 
in GaN these times are comparable~\cite{Brimont2009:APL}, which modifies not only the used REs,
but also the microscopic gain calculations since holes are also spin polarizaed~\cite{FariaJunior2017:PRB}. 
Such considerations are also relevant for bulk GaN-based spin-polariton lasers, where the spin polarization 
of holes was neglected~\cite{Bhattacharya2017:PRL}. There, with the edge emission, an in-plane magnetization of
FeCo/MgO spin injector requires no magnetic field for room-temperature operation.
  
An alternative path to electrical operation of spin-lasers  
at room temperature is sought with external-cavity lasers (VECSELs, noted in Section~1)~\cite{Frougier2015:OE}, 
shown in Fig.~\ref{fig:outlook}(c).  They could enable depositing a thin-film ferromagnet just 100-200 nm 
away from the active region, sufficiently close to attain a considerable spin polarization of carriers in the gain region 
at room temperature. With complementary properties to better-studied spin-VCSELs, such spin-VECSELs could
offer higher power and reduced noise. Recent progress in modeling VECSELs using a vectorial model~\cite{Alouini2018:OE}
could provide important guidance how the design of optical anisotropies enables an improved polarization dynamics,
reflecting their behavior as overdamped harmonic oscillators.

For both VECSEL and VCSEL geometries it is desirable to implement a spin injector with perpendicular anisotropy 
allowing operation in remanence~\cite{Sinsarp2007:JJAP, Hovel2008:APLa,Zarpellon2012:PRB}. 
An independent progress in spintronics to store and sense information using magnets with a perpendicular anisotropy~\cite{Tsymbal:2019}
could then also be directly beneficial for spin-lasers. Electrical spin injection usually relies on magnetic thin films with in-plane 
anisotropy requiring a large applied magnetic field to achieve an out-of-plane magnetization and the projection of injected 
spin compatible with the carrier recombination of circularly polarized light in surface-emitting lasers.  

\begin{figure}[t]
\centering
\includegraphics[width=10.8cm]{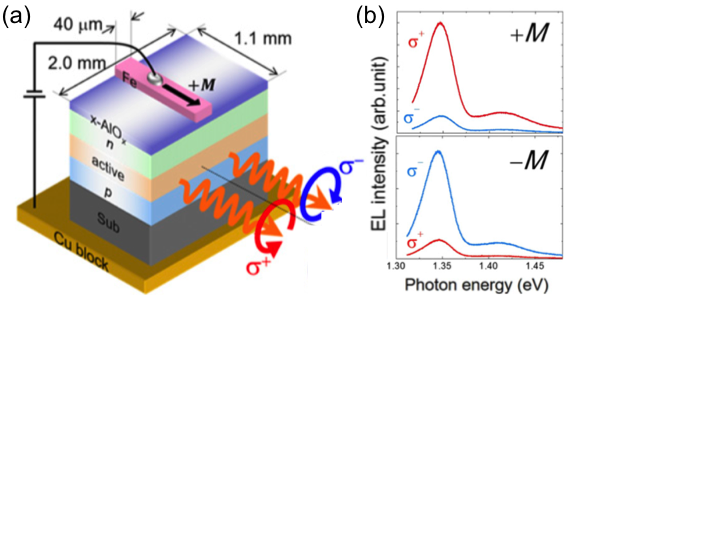}
\vspace{-3.8cm}
\caption{(a) A setup for the spin-LED in which the helicity of the room-temperature electroluminescence, EL, can be reversed by changing the 
direction of the magnetization, {\bf M}, in the Fe contact.  (b) Helicity-resolved EL spectra for {$+$ \bf M} as shown (a) in the upper panel and 
for {$-$\bf M} in the lower panel. These results show nearly completely circularly
polarized light at the maximum EL emission. From Ref.~\cite{Nishizawa2017:PNAS}.  
}
\label{fig:SLED}
\end{figure}

For spin modulation of lasers a steady-state electrical spin injection alone may not be enough and 
it is useful to recognize related advances in spintronics pertaining to magnetization dynamics, and variety of phenomena:
from spin-transfer and spin-orbit torques, to spin pumping~\cite{Tsymbal:2019}. The progress in fast magnetization reversal~\cite{Garzon2008:PRB,Kimel2005:N,Kirilyuk:2019} and spin-LEDs showing electrical helicity 
switching~\cite{Nishizawa2014:APL,Nishizawa2017:PNAS,Nishizawa2018:APE,Munekata2020:P}, see Fig.~\ref{fig:SLED}, 
could stimulate all-electrical schemes for spin modulation in lasers, to realize enhanced bandwidth discussed in Sections~3 and 5.
Remarkably, unlike the  
expected maximum of 50 \% circular polarization~\cite{Zutic2004:RMP} for a bulk III-V semiconductor
used in Fig.~\ref{fig:SLED}, such a spin-LED demonstrates room-temperature $P_C=95$ \% in remanence! 
Such helicity control is promising for secure communications and biomedical applications~\cite{Nishizawa2017:PNAS,Nishizawa2020:JJAP}
To understand this behavior and using it  
in electrically-operated spin-lasers, it is important to develop more general models of spin-transport and
accurate gain calculations, 
which would benefit from recent advances in ${\bf k \cdot p}$ methods~\cite{To2019:T}.  

Another direction to realize novel spin-lasers could build on a huge interest in two-dimensional materials and van der Waals heterostructures
which are suitable for low-power spintronic applications~\cite{Lin2019:NE}. Following our suggestion for atomically-thin spin-lasers based on a gain region made of  transition metal dichalcogenides ~\cite{Lee2014:APL,Lindemann2019:N},
we anticipate that their room-temperature electrical operation could be realized through magnetic proximity effects~\cite{Zutic2019:MT}.
A leaking magnetism from a nearby magnet would create proximity-induced spin-splitting~\cite{Zutic2019:MT,Zhao2017:NN,Zhong2017:SA} 
that could be changed by electric gating or altering the direction of magnetization~\cite{Lazic2016:PRB,Xu2018:NC}. 
Moreover, magnetic proximity effects in transition metal dichalcogenides could transform the optical selection
rules and tightly-bound electron-hole pairs~\cite{Scharf20017:PRL}, providing an intriguing opportunity for tunable spin-lasers.
    
\section*{Acknowledgments}
This work has been supported by the NSF ECCS-1810266, US ONR N000141712793, 
German Research Foundation, grants No. 250699912 and 392782903, and SFB 1277,
FAPESP grants No. 2012/05618-0 and 2013/23393-8, Alexander von
Humboldt Foundation, Capes grants No. 99999.000420/2016-06 and
88881.068174/2014-01, CNPq grants No. 408916/2018-4 and 308806/2018-2.
We thank S. Bearden, G. Boeris, D. Cao, Y.-F. Chen, H. Dery, 
C. G\o thgen, N. Jung, R. Michalzik, T.  Pusch, and E. Wasner  for valuable discussions.

\bibliographystyle{apsrev4-1}

\begin{thebibliography}{99}
\bibitem{const}
This is not essential but simplifies our analytical results.
\end{thebibliography}

\end{document}